\documentclass[twocolumn]{aastex61}
\usepackage{natbib,twoopt}
\usepackage{color}

\shorttitle{The circumstellar environment of MWC\,137}
\shortauthors{Kraus et al.}

\begin{document}
\title{Resolving the circumstellar environment of the Galactic B[e] supergiant star MWC\,137 from large to small scales\footnote{Based on observations collected with 
       1) the ESO VLT Paranal Observatory under programs 094.D-0637(B) 
          and 097.D-0033(A), 
       2) the MPG 2.2m-telescope at La Silla Observatory, Chile, under programs 096.A-9030(A) and  
          096.A-9039(A), 
       3) the Gemini Observatory, which is operated by the Association of Universities
          for Research in Astronomy, Inc., under a cooperative agreement with the
          NSF on behalf of the Gemini partnership: the National Science
          Foundation (United States), the Science and Technology Facilities
          Council (United Kingdom), the National Research Council (Canada),
          CONICYT (Chile), the Australian Research Council (Australia),
          Minist\'{e}rio da Ci\^{e}ncia, Tecnologia e Inova\c{c}\~{a}o (Brazil)
          and Ministerio de Ciencia, Tecnolog\'{i}a e Innovaci\'{o}n Productiva
          (Argentina), under the program GN-2013B-Q-11, 
       4) the Nordic Optical Telescope, operated by the Nordic Optical 
          Telescope Scientific Association at the Observatorio del Roque de los Muchachos, La Palma, 
          Spain, of the Instituto de Astrofisica de Canarias.
       5) the APEX telescope under the program CHILE-9711B-2016. APEX is a collaboration between the 
          Max-Planck-Institut f{\"u}r      
          Radioastronomie, the European Southern Observatory, and the Onsala Observatory.       
          }}

\correspondingauthor{Michaela Kraus}
\email{michaela.kraus@asu.cas.cz}

\author{Michaela Kraus}
\affiliation{Astronomick\'y \'ustav, Akademie v\v{e}d \v{C}esk\'e republiky, v.v.i., Fri\v{c}ova 298, 
251\,65 Ond\v{r}ejov, Czech Republic}
\affiliation{Tartu Observatory, 61602 T{\~o}ravere, Tartumaa, Estonia}

\author{Tiina Liimets}
\affiliation{Tartu Observatory, 61602 T{\~o}ravere, Tartumaa, Estonia}
\affiliation{Institute of Physics, University of Tartu, Ravila 14c, 50411, Tartu, Estonia}

\author{Cristina E. Cappa}
\affiliation{Instituto Argentino de Radioastronom{\'{\i}}a, CONICET, CCT-La Plata, C.C.5., 1894, Villa 
Elisa, Argentina}
\affiliation{Facultad de Ciencias
Astron\'omicas y Geof\'isicas, Universidad Nacional de La Plata, Paseo del Bosque s/n, 1900, La Plata, 
Argentina}

\author{Lydia S. Cidale}
\affiliation{Facultad de Ciencias Astron\'omicas y 
Geof\'isicas, Universidad Nacional de La Plata, Paseo del Bosque s/n, 1900, La Plata, Argentina}
\affiliation{Instituto de Astrof\'isica de La Plata, CCT La Plata, CONICET-UNLP, 
Paseo del Bosque s/n, 1900, La Plata, Argentina}
\affiliation{Instituto de F\'{i}sica y Astronom\'{i}a, Facultad de Ciencias, Universidad de 
Valpara\'{i}so, Av. Gran Breta\~na 1111, Casilla 5030, Valpara\'{i}so, Chile}

\author{Dieter H. Nickeler}
\affiliation{Astronomick\'y \'ustav, Akademie v\v{e}d \v{C}esk\'e republiky, v.v.i., Fri\v{c}ova 298, 
251\,65 Ond\v{r}ejov, Czech Republic}

\author{Nicolas U. Duronea}
\affiliation{Instituto Argentino de Radioastronom{\'{\i}}a, CONICET, CCT-La Plata, C.C.5., 1894, Villa 
Elisa, Argentina}

\author{Maria L. Arias}
\affiliation{Facultad de Ciencias Astron\'omicas y 
Geof\'isicas, Universidad Nacional de La Plata, Paseo del Bosque s/n, 1900, La Plata, Argentina}
\affiliation{Instituto de Astrof\'isica de La Plata, CCT La Plata, CONICET-UNLP, Paseo del Bosque s/n, 
1900, La Plata, Argentina}

\author{Diah S. Gunawan}
\affiliation{Instituto de F\'{i}sica y Astronom\'{i}a, Facultad de Ciencias, Universidad de 
Valpara\'{i}so, Av. Gran Breta\~na 1111, Casilla 5030, Valpara\'{i}so, Chile}

\author{Mary E. Oksala}
\affiliation{California Lutheran University, Department of Physics, Thousand Oaks, California 91360, 
USA}
\affiliation{LESIA, Observatoire de Paris, PSL Research University, CNRS, Sorbonne Universit\'es, UPMC Univ. Paris 06, Univ. Paris Diderot, Sorbonne Paris Cit\'e, 5 place Jules Janssen, F-92195 Meudon, France}

\author{Marcelo Borges Fernandes}
\affiliation{Observat\'orio Nacional, Rua General Jos\'e Cristino 77, 20921-400 S\~ao Cristov\~ao, Rio 
de Janeiro, Brazil}

\author{Grigoris Maravelias}
\affiliation{Instituto de F\'{i}sica y Astronom\'{i}a, Facultad de Ciencias, Universidad de 
Valpara\'{i}so, Av. Gran Breta\~na 1111, Casilla 5030, Valpara\'{i}so, Chile}

\author{Michel Cur\'{e}}
\affiliation{Instituto de F\'{i}sica y Astronom\'{i}a, Facultad de Ciencias, Universidad de 
Valpara\'{i}so, Av. Gran Breta\~na 1111, Casilla 5030, Valpara\'{i}so, Chile}

\author{Miguel Santander-Garc\'{i}a}
\affiliation{Observatorio Astron\'omico Nacional (IGN), C/ Alfonso XII 3, E-28014, Madrid, Spain}

\begin{abstract}
The Galactic object MWC\,137 was suggested to belong to the group of B[e] supergiants. However, with
its large-scale optical bipolar ring nebula and the high velocity jet and knots, it is a rather 
atypical representative of this class. We performed multi-wavelength observations spreading from the 
optical to the radio regime. Based on optical imaging and long-slit spectroscopic data we found that 
the northern parts of the large-scale nebula are predominantly blue-shifted, while the 
southern regions appear mostly red-shifted. We developed a geometrical model consisting of two 
double-cones. While various observational features can be approximated with such a scenario, the 
observed velocity pattern is more complex. Using near-infrared integral-field unit spectroscopy we 
studied the hot molecular gas in the close vicinity of the star. The emission from the hot CO gas 
arises in a small-scale disk revolving around the star on Keplerian orbits. While the disk itself 
cannot be spatially resolved, its emission is reflected by dust arranged in arc-like structures 
and clumps surrounding MWC\,137 on small scales. In the radio regime we mapped the cold molecular gas 
in the outskirts of the optical nebula. We found that large amounts of cool molecular gas and warm dust 
embrace the optical nebula in the east, south and west. No cold gas or dust were detected in the north 
and north-western regions. Despite the new insights on the nebula kinematics gained from our studies, the real formation scenario of the large-scale nebula remains an open issue.
\end{abstract}


\section{Introduction}

The enigmatic object MWC\,137 (V1308\,Ori) belongs to the group of Galactic B[e] stars.
It is surrounded by the optical nebula Sh 2-266 and located in the center of a cluster.
The evolutionary state of MWC\,137 has long been debated. Suggestions ranged from pre-main sequence 
\citep{1992ApJ...397..613H, 1992ApJ...398..254B, 1994A&AS..104..315T, 2016A&A...590A..98H} to post-
main-sequence, spreading over a large luminosity range \citep{1972ApJ...174..401H, 1984A&AS...55..109F, 
1998MNRAS.298..185E, 2013A&A...558A..17O}.

Significant progress in the star's classification was achieved by \citet{2015AJ....149...13M}, who
modeled the emission from hot $^{13}$CO gas in the vicinity of MWC\,137 that was first detected by 
\citet{2013A&A...558A..17O}. The presence of measurable 
amounts of $^{13}$CO implies a significant enrichment of the circumstellar material in $^{13}$C 
\citep{2009A&A...494..253K}. As stellar evolution models show \citep[e.g.][]{2012A&A...537A.146E}, this 
isotope is processed inside the star and via mixing processes transported to the surface, from which
it is liberated into the environment by mass loss events. With the discovery of hot, circumstellar 
$^{13}$CO emission, a pre-main sequence evolutionary phase of MWC\,137 could finally be excluded. 

The evolved nature of MWC\,137 is further consolidated by the studies of \citet{2016A&A...585A..81M}.
These authors investigated the whole cluster and determined a cluster age of $>$ 3 Myr. With a mass
of 10-15\,M$_{\sun}$ for MWC\,137, this object has clearly evolved off the main sequence.

Moreover, \citet{2016A&A...585A..81M} discovered a jet with several individual knots emanating from 
MWC\,137 with high velocities. Estimates of the age of these knots revealed that the jet must be much
younger than the large-scale optical nebula. The position angle of the jet is aligned with the 
polarization angle and hence perpendicular to the circumstellar disk traced on small scales by 
intrinsic polarization in H$\alpha$ \citep{1999MNRAS.305..166O}. Further confirmation for a rotating 
circumstellar disk was provided by the rotationally broadened CO band head emission reported by 
\citet{2015AJ....149...13M}. 

The large-scale structure seen in H$\alpha$ images \citep[e.g.,][]{2008A&A...477..193M} led
\citet{1998MNRAS.298..185E} to suggest that Sh\,2-266 could be a ring nebula produced by the 
interaction of the stellar wind with the interstellar medium. On the other hand, as the morphology of 
Sh\,2-266 is reminiscent of bipolar ring nebulae detected around two early B-type supergiants, Sher 25
\citep{1997ApJ...475L..45B} and SBW1 \citep{2007AJ....134..846S}, \citet{2015AJ....149...13M} proposed 
that the nebula material might have been ejected during the blue supergiant phase so that MWC\,137 
might be transiting from a B[e] supergiant into a blue supergiant with a bipolar ring nebula.

We carried out an observational campaign combining data from various wavelength regimes (optical, 
infrared, and radio) and on different spatial scales. These data are aimed to investigate in detail the 
structure and kinematics of the environment of MWC\,137 on both the large and small scale.

\section{Observations and data reduction}

\subsection{Optical imaging and spectroscopy}
Long slit optical spectra and imaging data were obtained on 2016 November 8 with the Nordic Optical 
Telescope (NOT) using the Andalucia Faint Object Spectrograph and Camera (ALFOSC).
For the imaging one long (600 sec) and one short (10 sec) exposure were obtained using the broad band 
H$\alpha$ filter (Halp\_658\_18 No. 22), which includes also the [N\,{\sc ii}] $\lambda\lambda$ 
6548,6583 doublet. The short exposure was used to fix the exact stellar coordinates, while the long 
exposure resolved the spatial structure of the nebula and served as template to position the slits. 
ALFOSC's field of view (FOV) is $6\farcm 4\times6\farcm 4$ and the pixel scale is $0\farcs 21$ 
pix$^{-1}$.

For the long-slit observations we utilized Grism~\#17 with a slit width of $0\farcs 5$, providing a 
spectral coverage from 6315~\AA\ to 6760~\AA\ and a spectral reciprocal dispersion of 0.29~\AA\ 
pix$^{-1}$. This spectral range was chosen to trace H$\alpha$ and the strongest nebular lines: 
[N\,{\sc ii}] $\lambda\lambda$ 6548,6583 and [S\,{\sc ii}] $\lambda\lambda$ 6716,6731. Due to poor 
weather conditions, only three slit positions could be observed. Two of them were centered on the star 
and had position angles (PA) of 35$\degr$ and 298$\degr$ (see details in Sect.\,\ref{kinematics} and 
Fig.\,\ref{fig:rv}). For both an integration time of 30 min was 
used in a single exposure. The third position was chosen slightly off the central star with two field 
stars as reference points and with a PA of 341.1$\degr$. It was aligned such that individual knots, in 
particular knot c of the jet were covered. For this position, two exposures of 30 min each were 
acquired and co-added.

The star was also observed with medium resolution using the Coud\'{e} spectrograph 
\citep{2002PAICz..90....1S} attached to the Perek 2-m telescope at Ond\v{r}ejov Observatory. Three 
spectra were collected, two in the H$\alpha$ region (6250--6760\,\AA) on 2011 October 1 and 2013 
October 5, and one in the red region 
(6990--7500\,\AA) on 2013 October 5. The observations were carried out with the 830.77 lines\,mm$^{-1}$ 
grating and a SITe $2030\times 800$ CCD for the spectra taken before June 2013, and a  
PyLoN $2048\times 512$BX CCD for those taken thereafter. With a slit width set at 0\farcs 6, the 
spectral resolution is $R\simeq$ 13\,000 in the H$\alpha$ region and $R\simeq$ 15\,000 in the 
red region.

In addition, we acquired high-resolution spectra on 2015 December 5 and 2016 February 28 with the 
Fiber-fed Extended Range Optical Spectrograph \citep[FEROS][]{1999Msngr..95....8K} attached to the MPG 
2.2-m telescope at the European Southern Observatory in La Silla (Chile). FEROS is a bench-mounted 
echelle spectrograph with fibers, which cover a sky area of 2$\arcsec$ of diameter. The wavelength 
coverage ranges from 3600\,\AA \ to 9200\,\AA, and the spectral resolution is R = 48\,000 (in the 
region around 6000 \AA).

To perform telluric corrections, spectra of a standard star, typically a rapidly rotating 
early-type star, were taken during the observing nights.

The data collected at the NOT and Ond\v{r}ejov were reduced (bias, flat fielding, wavelength 
calibration) using standard IRAF\footnote{IRAF is distributed by the National Optical Astronomy 
Observatory, which is operated by the Association of Universities for Research in Astronomy (AURA) 
under cooperative agreement with the National Science Foundation.} routines, whereas for the FEROS data 
the reduction pipeline was utilized. Telluric correction was performed for the FEROS and the 
Ond\v{r}ejov data, and all spectra were corrected for heliocentric velocity.

\subsection{Near-infrared spectroscopy}
MWC\,137 was observed on 2014 December 30 and 2016 March 19 with the Spectrograph for INtegral Field
Observation in the Near-Infrared \citep[SINFONI;][]{2003SPIE.4841.1548E, 2004Msngr.117...17B} on the 
ESO VLT UT4 8-m telescope. 
The observations were carried out using the $0\farcs 8\times 0\farcs 8$ FOV (25 mas plate scale) with 
the K-band grating, which provides a spectral resolution of $R=4500$ and a wavelength coverage of 
$1.95-2.45\,\mu$m. For proper sky subtraction, the observations were carried out in an ABBA nod 
pattern with the B position taken 1\arcmin \ west of the central star. 
Data reduction was performed with the ESO SINFONI pipeline (version 2.7.0). Raw frames were 
corrected for bad pixels, flat fields, non-linearity, and were wavelength calibrated. 

In addition, the star was observed on 2013 December 14 with the Gemini Near-Infrared Spectrograph 
(GNIRS) mounted on the Gemini North telescope in the L-band. We used the long camera with grating 110.5 
l\,mm$^{-1}$ and $0\farcs 10$ slit which we centered on Br$\alpha$. The achieved spectral resolution is 
$R\simeq 19\,000$ with a wavelength coverage of $\sim 0.1 \mu$m. To facilitate sky subtraction, 
the observations were carried out in an ABBA nod pattern along the slit with an offset of 6\arcsec \ 
between the A and B positions.
The data were reduced using standard IRAF routines.

In both cases, a telluric standard star was observed at similar airmass. Telluric and heliocentric 
velocity corrections of the infrared spectra were performed with standard IRAF tasks.

\subsection{Radio data}

Radio observations were taken on 2016 March 21 and 22, July 30 and 31, and August 1st with the Atacama Pathfinder EXperiment 
\citep[APEX;][]{2006A&A...454L..13G} located at Llano de Chajnantor (Chilean Andes). As front end for 
the observations, we used the APEX-2 receiver ($T_{\rm sys}$ = 300 K) of the  Swedish Heterodyne 
Facility Instrument \citep[SHeFI;][]{2008A&A...490.1157V} to perform observations of the molecular line 
$^{12}$CO(3-2) at 345.79599 GHz and $^{13}$CO(3-2) at 330.58796 GHz, and the APEX-1 receiver (150 K) to observe the SiO(5-4) and CS(5-4) 
molecular lines at 217.104980 and 244.935644 GHz, respectively. The observations were carried out 
within an area of 3\arcmin $\times$3\arcmin\ centered at the position of the B[e] star (\hbox{RA, 
Dec.(J2000)} = 06$^h$18$^m$46$^s$, 15\degr16\arcmin50\arcsec)

Observations were made using the on-the-fly (OTF) mode with two orthogonal scan directions along RA and 
Dec. and a space between dumps in the scanning direction of 9\arcsec.  Calibration and pointing 
were performed using IRC+10216, RAFGL\,865, OMC1, IK-TAU, R Dor, o Ceti, and CRL\,618.  The intensity 
calibration has an uncertainty of  10\%.

The backend  for the observations was the FFT spectrometer consisting of 4096 channels, with a total 
bandwidth of 1000\,km\,s$^{-1}$ and a velocity resolution of 0.2\,km\,s$^{-1}$.  Atmospheric 
attenuation correction was done by skydips, and the output intensity scale given by the system is 
$T_{\rm A}$.  The observed intensities were converted to the main-beam brightness-temperature scale by 
$T_{\rm mb}$ = $T_{\rm A}$/$\eta_{\rm mb}$, where   $\eta_{\rm mb}$ is the main beam efficiency. For 
the SHeFI/APEX-1 and  SHeFI/APEX-2 receiver the adopted value is $\eta_{\rm mb}$ = 0.73 
\citep{2008A&A...490.1157V}.  The half-power beam-width of the telescope is  $\sim$  21\arcsec\ for the 
CO lines and 30\arcsec\ for the SiO and CS lines. The off-source position free of  CO emission was 
located at  \hbox{RA, Dec.(J2000)} = 6$^h$34$^m$33.8$^s$, +16\degr 52\arcmin 48\farcs 9.

The spectra were reduced using the Continuum and Line Analysis Single-dish Software (CLASS90) programme 
of the IRAM GILDAS software package\footnote{http://www.iram.fr/IRAMFR/GILDAS}. A linear baseline 
fitting was applied to the data. The rms noise of the profiles after baseline subtraction and 
calibration is $\sim$ 0.3 K for $^{12}$CO(3-2) and 0.2 K for $^{13}$CO(3-2) per spectral channel. The Astronomical Image Processing 
System (AIPS) package  was used to perform the analysis.

\hfill

\subsection{Complementary data}

We searched for additional information provided in different archives and retrieved images in the 
infrared and in the radio continuum from public surveys that complement our own data sets.

The region around MWC\,137 was observed in the near- and mid-infrared with the Wide-field Infrared 
Survey Explorer 
\citep[WISE;][]{2010AJ....140.1868W} satellite. The four bands are centered on 3.4, 4.6, 12.0, and 
22.0 $\mu$m and have angular resolutions of 6\farcs 1, 6\farcs 4, 6\farcs 5, and 12\farcs 0, 
respectively. Moreover, the Galactic Legacy Infrared Mid-Plane Survey 
Extraordinaire \citep[GLIMPSE;][]{2001AAS...198.2504C, 2003PASP..115..953B} imaged MWC\,137 using the 
two Infrared Array Camera \citep[IRAC;][]{2004ApJS..154...10F} bandpasses at 3.6 and 4.5 $\mu$m on the 
{\em Spitzer Space Telescope} \citep{2004ApJS..154....1W}. The angular resolutions are less than 
2\arcsec\ in all bands. 

In the far-infrared (FIR) regime the {\em Herschel Space Observatory}\footnote{{\em Herschel} is an ESA 
space observatory with science instruments  provided by European-led Principal Investigator consortia 
and with important participation from NASA (http://www.cosmos.esa.int/web/herschel/science-archive)} 
imaged MWC\,137 at 70 and 160\,$\mu$m with the Photodetector Array Camera and Spectrometer 
\citep[PACS;][]{2010A&A...518L...2P} and at 250, 350, and 500\,$\mu$m with the Spectral and 
Photometric Imaging REceiver \citep[SPIRE;][]{2010A&A...518L...3G}. These data were acquired during 
the Herschel Infrared GALactic plane survey \citep[Hi-GAL;][]{2010A&A...518L.100M} key program. 
The angular resolutions at 70, 160, 250, 350 and 500\,$\mu$m are 8\farcs 5, 13\farcs 5, 18\arcsec, 
25\arcsec, and 36\arcsec, respectively.

All infrared images were obtained from the Infrared Science Archive\footnote{IRSA, 
http://irsa.ipac.caltech.edu} provided by the NASA Infrared Processing and Analysis Center (IPAC). 
The retrieved Herschel images were processed at level 2.5 (PACS) and 3 (SPIRE).

In addition, we retrieved radio continuum data at 1420 MHz that were taken during the NRAO Very Large 
Array (VLA) Sky Survey \citep[NVSS;][]{1998AJ....115.1693C}. The images have a spatial resolution of 
45\arcsec\ and a rms noise of about 0.45\,mJy\,beam$^{-1}$ (Stokes I).

\section{Results}

\subsection{Star and optical nebula}

\begin{figure}
\begin{center}
\includegraphics[width=\hsize,angle=0]{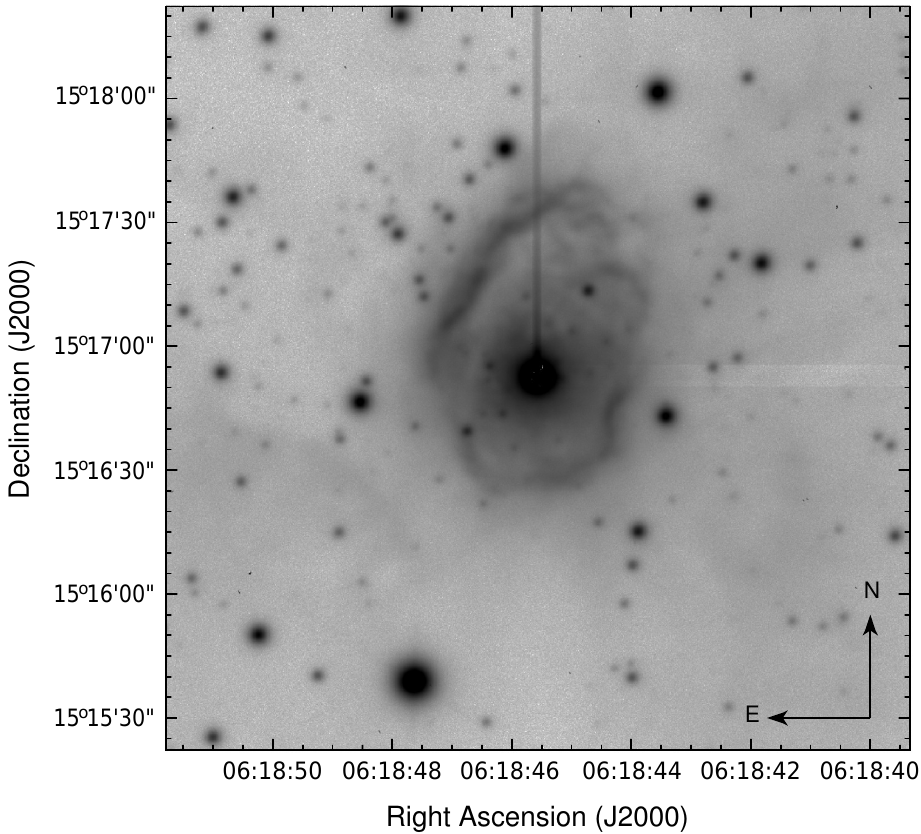}
\caption{ALFOSC H$\alpha$ image of the nebula structure around MWC\,137. The FOV is 
$3\arcmin\times3\arcmin$. The dark line from the star to the north and the bright wide band to the west  
are instrumental artifacts.}
\label{fig:alfosc} 
\end{center}
\end{figure}

The structure of the optical nebula around MWC\,137 is resolved in the long exposure ALFOSC narrow band 
image in the H$\alpha$+[N\,{\sc ii}] filter (Figure \ref{fig:alfosc}). The nebula has an oval shape 
with arc-like structures along its outer boundary and a size of $80\arcsec\times60\arcsec$, similar to 
what was resolved in earlier observations \citep{2008A&A...477..193M, 2016A&A...585A..81M}. The
ionized nebula displays an asymmetry with respect to the stellar position. It is more extended to the 
north. We also note diffuse emission at larger distances, in particular south-east and south-west
of the main nebula.

\begin{figure}
\begin{center}
\includegraphics[width=\hsize,angle=0]{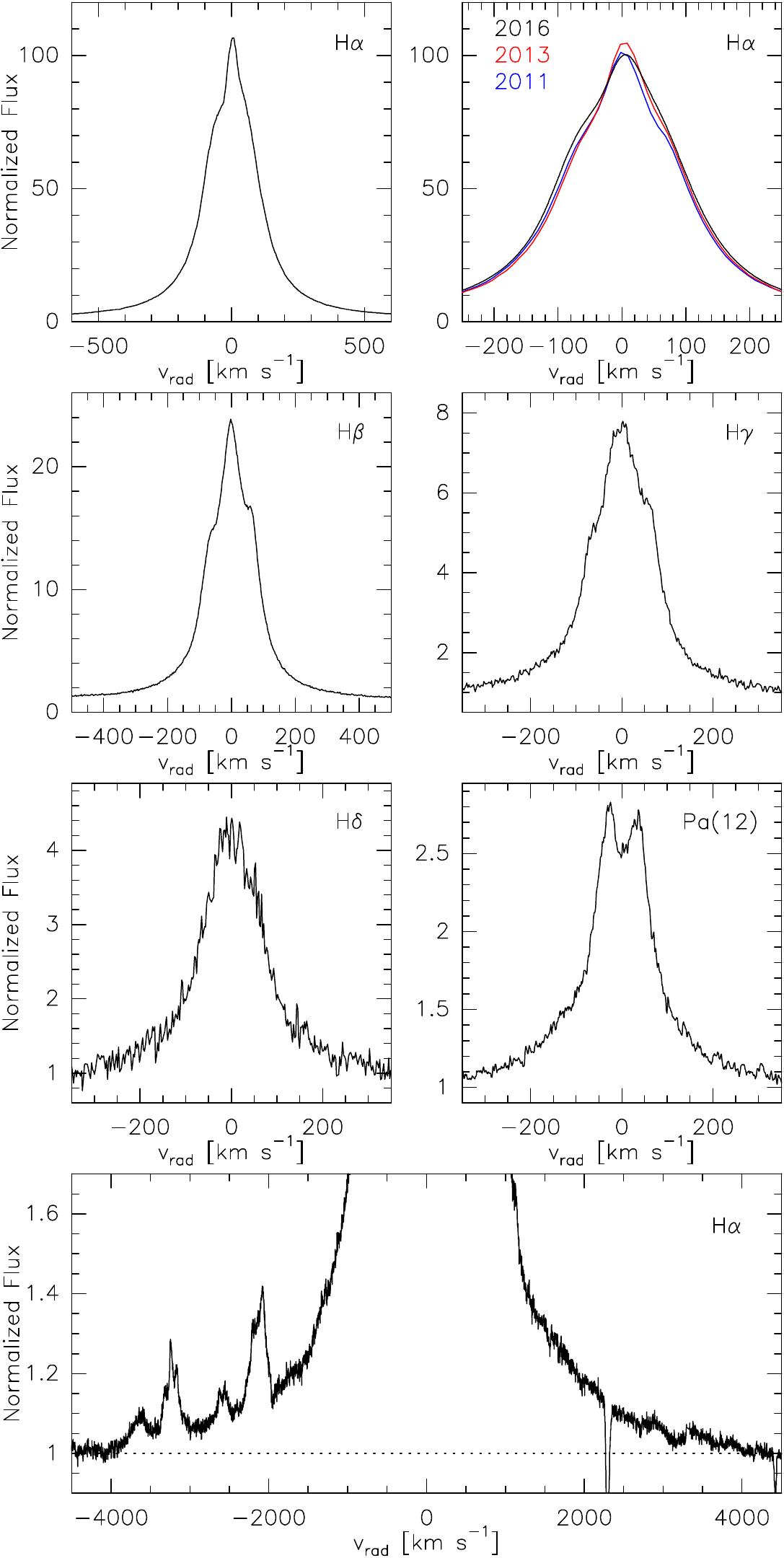}
\caption{Profiles of the hydrogen lines of the lower Balmer series and of Pa(12). The top right panel
shows the H$\alpha$ line obtained in Ond\v{r}ejov in 2011 (blue) and 2013 (red) 
compared to the profile in the FEROS spectrum from 2016. The latter was convolved to the same low 
resolution. The broad emission wings of H$\alpha$ are depicted in the bottom panel.}
\label{fig:Balm} 
\end{center}
\end{figure}

The Ond\v{r}ejov and FEROS spectra cover the innermost 0\farcs 6 and 2\arcsec, 
respectively, which contain the star and its closest environment. The spectra display no 
obvious photospheric absorption lines, but numerous emission lines from both permitted and forbidden
transitions of elements in different ionization stages, as was also reported in earlier works of 
\citet{2003A&A...408..257Z} and \citet{2004AJ....127.1682H}. Absorption features in the spectra 
are due to diffuse interstellar bands and cool material in foreground clouds traceable, e.g., via
the absorption in the K\,{\sc i} 7699\,\AA \ line as previously mentioned by 
\citet{2016A&A...590A..98H}.

To determine the systemic velocity of MWC\,137, we measured the central velocities of a sample 
of symmetric forbidden emission lines. We obtained an average radial velocity of $42.0\pm 
0.6$\,km\,s$^{-1}$ which we used to correct all optical and infrared spectra.

Comparison of our spectra from different epochs revealed no obvious variability in the emission lines, 
neither in the Ond\v{r}ejov spectra that have a time difference of about two years, nor in the 
high-resolution FEROS spectra, which were taken about three months 
apart. Only a slight change in the profile of the H$\alpha$ line was noted in the Ond\v{r}ejov spectra. 
To compare with the FEROS data, we convolved the latter to the resolution of 13\,000 of the 
Ond\v{r}ejov spectra and confirm the earlier 
findings of \citet{2003A&A...408..257Z} that H$\alpha$ is slightly changing over the years, as is shown 
in the upper right panel of Figure \ref{fig:Balm}. As the FEROS spectrum from 2016 
has the highest quality, all following line profile figures refer to these data.

Most prominent are the emission lines from the Balmer series (H$\alpha$ to H$\delta$, see Figure 
\ref{fig:Balm}). The profiles of H$\alpha$ to H$\gamma$ appear triple-peaked. The wings of H$\alpha$ 
are also remarkable which extend to velocities of 3500--4000\,km\,s$^{-1}$ (bottom panel 
of Figure \ref{fig:Balm}). However, this high value is not a dynamic velocity. Instead, as for all B[e] 
supergiants, these wings are created by electron scattering. \citet{2003A&A...408..257Z} reported on 
much smaller H$\alpha$ wings with velocities reaching to only about half of our value. Such a 
drastic change in the wings might be caused by an increase the number of scatterers, i.e., free 
electrons, along the line of sight, possibly due to a higher mass-loss rate.  

\begin{figure}
\begin{center}
\includegraphics[width=\hsize,angle=0]{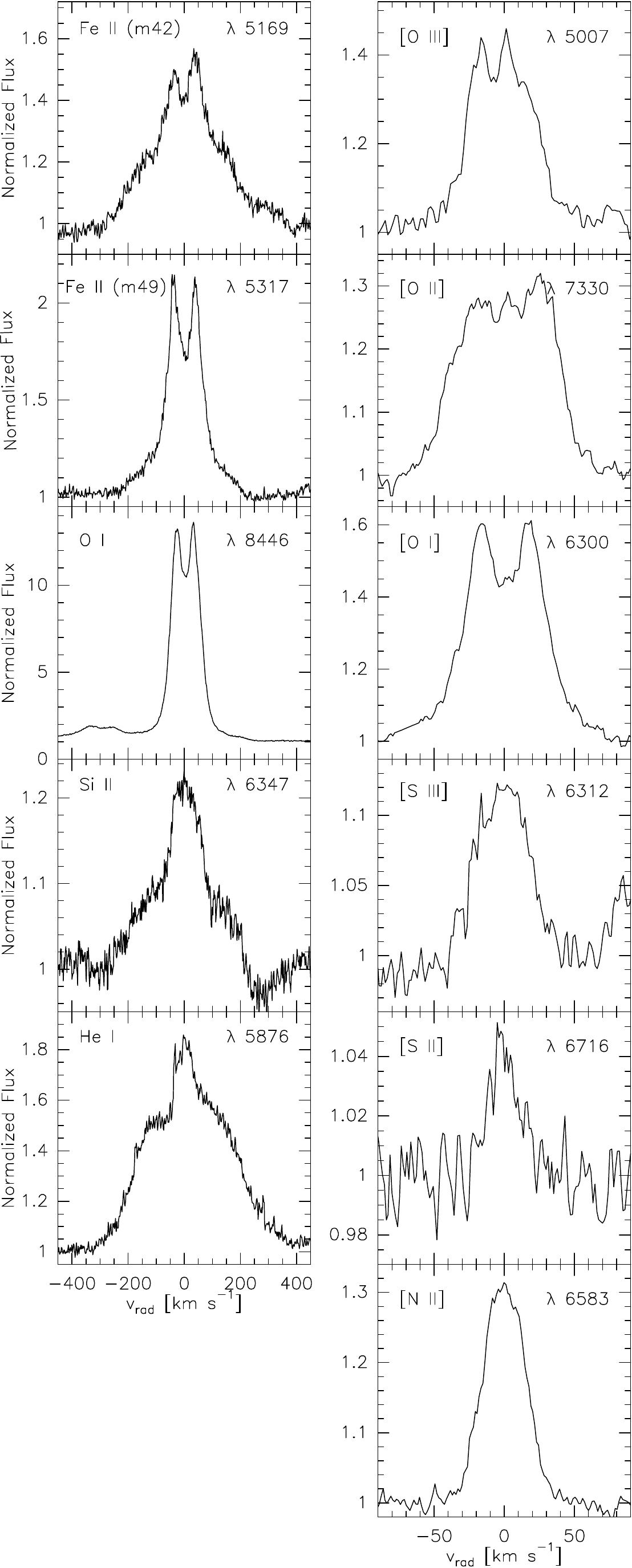}
\caption{Line profile variety in permitted (left) and forbidden (right) lines in different ionization states.}
\label{fig:profiles}
\end{center}
\end{figure}

Lines from the Paschen series are resolved up to Pa(30).
Their profiles appear composite consisting of a symmetric double-peaked component with a peak 
separation of 65--70\,km\,s$^{-1}$ on a weak but broad component with wings extending up to $\sim 
350$\,km\,s$^{-1}$, as shown exemplary for Pa(12) in Fig. \ref{fig:Balm}. The lines Pa(13), Pa(15), 
and Pa(16) are blended with the Ca{\sc ii} triplet lines. These calcium lines have double-peaked 
profiles, as was previously reported by \citet{1992ApJS...82..285H}. Also included in this wavelength 
range are the O{\sc i} $\lambda$ 8446 line and a few weak lines of Fe{\sc ii} and N{\sc i}, which all 
display double-peaked profiles. The full profile shape of the O{\sc i} $\lambda$ 8446 line is shown in 
the bottom left panel of Figure \ref{fig:profiles}. With a peak separation of $\sim$ 60\,km\,s$^{-1}$ 
this line appears slightly narrower than the Paschen double-peaks.

Similar profiles as for the Paschen lines are observed for the Fe{\sc ii} lines. Example profiles from 
multiplet 42 and 49 are shown in Figure \ref{fig:profiles}. The broad underlying component has clearly 
asymmetric line wings. 

The lines of He{\sc i} and Si{\sc ii} are alike the lower Balmer line profiles with a triple-peaked 
structure (Figure \ref{fig:profiles}). It is interesting to note that the triple-peaked 
profile in He{\sc i} $\lambda$ 5876 was also seen in 2002, while older observations from 1987 revealed 
that this line was in absorption \citep{2003A&A...408..257Z}. This turn from absorption to emission
is an indicator for higher densities along the line of sight, in agreement with the broader 
electron scattering wings of H$\alpha$.

In contrast to the broad permitted lines, the profiles of the forbidden emission lines are typically 
much narrower and single peaked. The only exceptions are the lines [O{\sc i}] $\lambda\lambda$ 
6300,6364 which display clear double-peaked profiles (see Figure \ref{fig:profiles} for an example) 
with peak 
separations of 32--33\,km\,s$^{-1}$. The lines of [O{\sc ii}] are typically blends. Their real profile 
shape is thus hidden. The co-existence of different ionization states of individual elements such as 
oxygen and sulfur is a clear hint for a non-spherical density distribution of the circumstellar 
material. In particular the presence of lines from O{\sc i} and O{\sc iii} is similar to the spectra
seen from the compact planetary nebula Hen 2-90 \citep{2005A&A...441..289K}. In this object, optical 
imaging revealed a very hot polar wind in which the emission from twice ionized elements such as 
O{\sc iii} arise, together with a cold equatorial disk from which the emission from neutral atoms and 
dust originate. 

A Keplerian rotating disk or ring scenario as the formation region of the double-peaked profiles of the 
[O{\sc i}] lines was proposed for the B[e] supergiants \citep{2010A&A...517A..30K} and seems now to 
be well established based on different (atomic and molecular) tracers \citep{2012MNRAS.423..284A, 
2012A&A...548A..72C, 2012A&A...543A..77W, 2015ApJ...800L..20K}. The double-peaked [O{\sc i}] lines 
might thus trace a compact disk around MWC\,137.

Interestingly, we observe neither [O{\sc i}] $\lambda$ 5577 nor [Ca{\sc ii}] $\lambda\lambda$ 
7291,7324. As these lines are typically associated with the denser regions of the circumstellar 
disks closer to the star than the [O{\sc i}] $\lambda\lambda$ 6300, 6364 line forming region 
\citep[see, e.g.,][]{2010A&A...517A..30K, 2012MNRAS.423..284A, 2016MNRAS.456.1424A, 
2017ASPC..508..213M}, the lack of observable emission in these lines suggests that the density in 
the cool, close-by environment of MWC\,137 might be lower than in other B[e] supergiants.

The wings of most of the forbidden emission lines extent to velocities of about 50\,km\,s$^{-1}$. Only 
the lines [N{\sc ii}] $\lambda\lambda$ 6548,6583 and [S{\sc ii}] $\lambda\lambda$ 6716,6731 are 
narrower ($\leq$ 30\,km\,s$^{-1}$). Moreover, these lines are very weak in our spectra, whereas they 
become very intense in the large-scale nebula.

\subsection{Kinematics of the large-scale optical nebula}\label{kinematics}

\begin{figure*}
\begin{center}
\includegraphics[width=8.5cm]{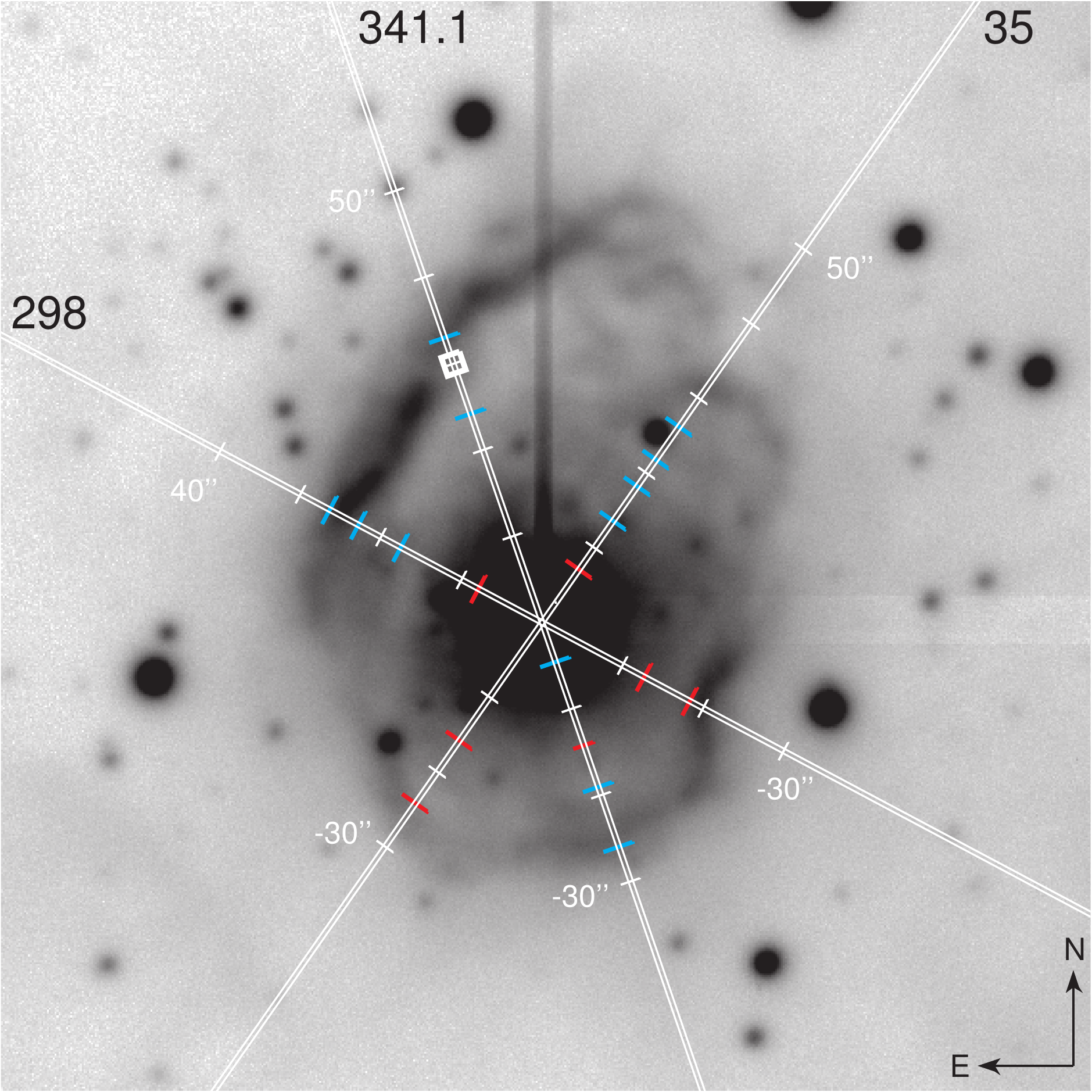}
\includegraphics[width=9.2cm]{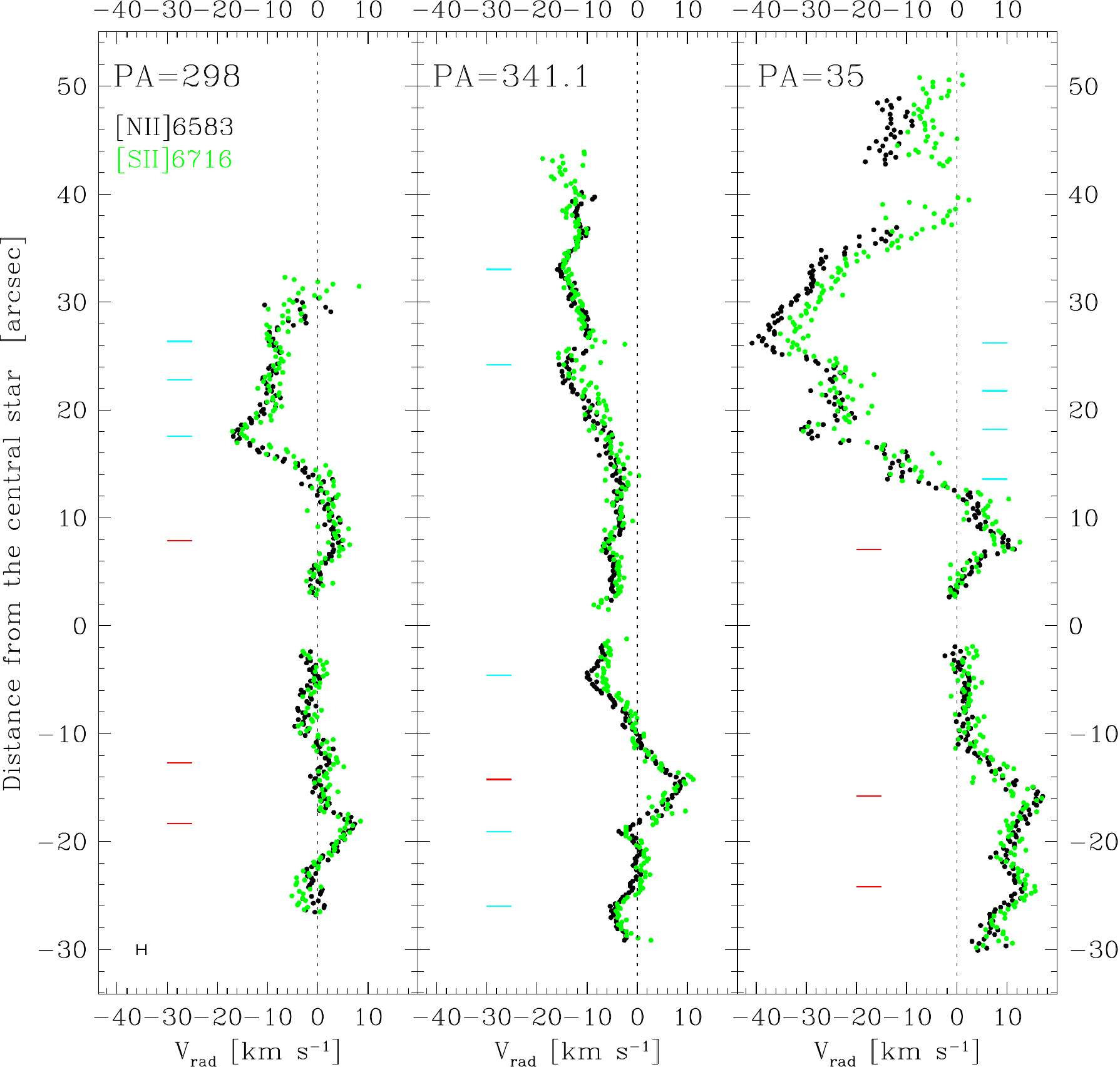}
\caption{\textit{Left.} Positions of the NOT slits (white) and their PA values (black numbers) overlaid 
on the zoomed (FOV of $2\arcmin\times2\arcmin$) ALFOSC H$\alpha$ image. A distance scale in steps of 
$10\arcsec$ is marked with white ticks along the slits in both directions from the central star. Red 
and cyan ticks mark positions of red- and blue-shifted RV extrema, respectively. PA = 341.1 is along 
the jet and the white square indicates knot c. \textit{Right.} Radial velocities of the [N\,{\sc ii}] 
6583~\AA\ (black) and [S\,{\sc ii}] 6716~\AA\ (green) emission lines. Positive distance from the 
central star on the image is towards the North, negative towards the South. Red and cyan ticks mark 
red- and blue-shifted velocity extrema determined for the [N\,{\sc ii}] 6583~\AA\ line. The average 
error bar for all the measurements is in the lower left corner.}
\label{fig:rv} 
\end{center}
\end{figure*}

To investigate the nebula kinematics, we make use of the long-slit spectra. The orientation and nebula 
coverage of the three available slit positions are presented in the left panel of Figure \ref{fig:rv}.

For the radial velocity (RV) measurements we used the two nebula lines [N\,{\sc ii}] $\lambda$ 6583 and 
[S\,{\sc ii}] $\lambda$ 6716, because they are the more intense ones from the doublets. These lines can 
be traced from the stellar position out to the edges of the optical nebula. As the images are strongly 
saturated in H$\alpha$, in particular in the vicinity of the central star's position, this line was 
excluded from the analysis. The measurements were done from the 2D spectra by fitting the line 
profiles, line by line with a single Gaussian. Occasionally, a cosmic ray fell on top of the line which 
could not be removed. Also, in some parts of the nebula the intensity was too weak for reliable 
measurements. These regions were thus excluded. The measurements were corrected for heliocentric and 
systemic velocities, and the resulting RV values along the three individual positionings of the slit
are shown in the right panel of Figure \ref{fig:rv}. The full information on the final RVs is available 
as an on-line table of the form shown in Table~\ref{tab:rv}. 

Both nebula lines display the same kinematic behavior, with RV values ranging 
from -41 to +18\,km\,s$^{-1}$. The errors of the individual measurements are relatively small, on 
the order of 0.9\,km\,s$^{-1}$ on average as shown by the error bar in the lower left corner of the
radial velocity plot. They were estimated based on the RMS of the wavelength calibration, whereas the
contribution from the fitting is negligible. Precise error values are included in Table~\ref{tab:rv}.

\begin{deluxetable}{lllll}
\tabletypesize{\scriptsize}
\tablecaption{Radial velocity measurements. \label{tab:rv} }
\tablewidth{0pt}
\tablehead{
\colhead{PA}   & \colhead{$\lambda_{\rm lab}$} & \colhead{RV} & Error RV  &\colhead{Dist}         \\
\colhead{$\degr$}  &  \colhead{\AA}   &  \colhead{km~s$^{-1}$}   & \colhead{km~s$^{-1}$} &\colhead{\arcsec}
}
\startdata
 35 & 6716.44 &   3.85 &  0.90 &-29.84 \\
 35 & 6716.44 &   5.23 &  0.90 &-29.63 \\
 35 & 6716.44 &  11.03 &  0.90 &-29.42 \\
 35 & 6716.44 &   8.27 &  0.90 &-29.21 \\
 35 & 6716.44 &   9.83 &  0.90 &-29.00 \\
... &         &        &       &       \\
\enddata
\tablecomments{The entire table is published only in the electronic
  edition of the article.  The first 5 lines are shown here for
  guidance regarding its form and content.}
\end{deluxetable}

The [S\,{\sc ii}] $\lambda$ 6716 line is considerably weaker than [N\,{\sc ii}] $\lambda$ 6583, 
resulting sometimes in a larger scatter of the measured values, in particular in regions of lower 
intensity as is the case at the edges of the nebula structure and in the northern region of PA = 35. 
Along this position angle the intensity in both lines basically disappears at distances 
38\arcsec--42\arcsec\ to the north so that the velocities could not be measured resulting in a gap in 
Figure \ref{fig:rv}.

\begin{deluxetable}{lrr}[!h]
\tablecaption{Extrema of the RV values of [N\,{\sc ii}] 6583~\AA\ marked in Figure \ref{fig:rv}. 
\label{tab:ext} }
\tablewidth{0pt}
\tablehead{
\colhead{PA}       & \colhead{RV}              & \colhead{Dist}         \\ 
\colhead{$\degr$}  &  \colhead{km~s$^{-1}$}    & \colhead{\arcsec}
}
\startdata
298   & 7.32    & -18.36\\
      & 2.27    &-12.69\\
      & 4.82    & 7.89 \\
      & -16.81  &17.55 \\
      & -10.98  &22.80 \\
      & -9.71   &26.37 \\
341.1 &  -5.41  &-25.99\\
      &  -3.68  &-19.06\\
      &  9.16   &-14.23\\
      &  -10.01 & -4.57\\
      &  -15.56 & 24.20\\
      &  -15.93 & 33.02\\      
35    & 13.35  &-24.20\\
      & 17.04  &-15.80\\
      & 11.39   &7.09  \\
      & -13.84  &13.60 \\
      & -31.00 &18.22 \\
      & -29.14 &21.79 \\
      & -40.79 &26.20 \\
\enddata
\end{deluxetable}

As a general trend we note that the emission from the northern nebula 
parts is predominantly blue-shifted, whereas the southern regions have mostly red-shifted emission. 
However, deviations from this general trend exist. For instance, the RV measurements for PA = 35 and 
PA = 298 display a red-shifted emission in the northern nebula part in the vicinity of the star 
(5\arcsec -- 15\arcsec). In addition, PA = 341.1 displays three southern regions with 
blue-shifted emission. In fact, we observe a series of extrema in the RV measurements along each slit 
position. They are marked by cyan and red ticks in both panels of Figure \ref{fig:rv} and represent
maxima in blue- and red-shifted velocities, respectively. These values are also listed in 
Table~\ref{tab:ext}.   

Most remarkable is the RV variation along PA = 35. It displays the highest amplitudes in both 
the blue- and red-shifted emission while the slit passes through nebula regions which appear less 
intense. The blue-shifted emission seen towards the north contains a series of pronounced maxima. 
The image of the nebula also shows that this region seems to have generally more structure. 
However, there is another star \citep[identified as B3-5 by][]{2016A&A...585A..81M} in the close 
vicinity of PA = 35 at a distance of $\sim$ 25\arcsec, which might influence the kinematics of the 
nebula.  

Close inspection of the positions of the major velocity extrema along all three slits in comparison 
with the intensity structure of the nebula reveals that they typically precede intensity
accumulations.

\subsection{Jet} 

\begin{figure}[!h]
\includegraphics[width=\hsize,angle=0]{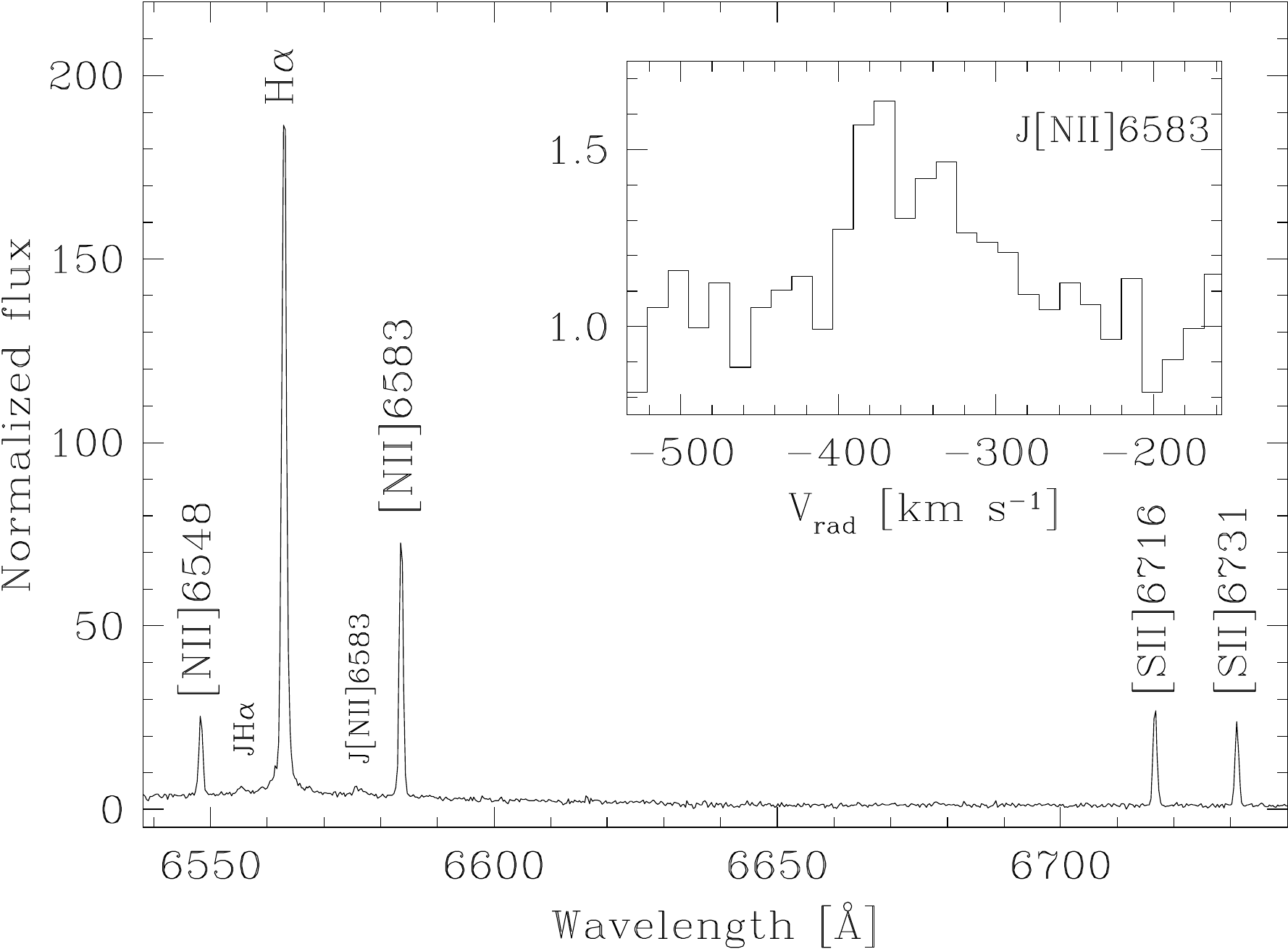}
\caption{ALFOSC spectrum of the nebula emission and the emission of the jet component c (denoted 
as J\,H$\alpha$ and J\,[N\,{\sc ii}]\,6583). The J\,[N\,{\sc ii}]\,6583 line of knot c in the 
insert indicates sub-structure over a large velocity range that was previously not reported.}
\label{fig:jetspec}
\end{figure}

The PA=341.1 was positioned such that it covered parts of the jet discovered by 
\citet{2016A&A...585A..81M}. As this jet appears slightly tilted (see their Figure 8) and the slit is 
rather narrow, only one clear signal from the jet is seen in our data: the emission from knot c. This 
feature is located at a distance of 30\arcsec \ from the central star and is marked by a white box in 
Figure \ref{fig:rv}. On the 2D image, knot c has a slightly elongated shape spreading over 
about $10\times10$\,pix. This size corresponds to $2\farcs 1$. 
We observe a clear inclination of the feature on the 2D image, meaning that the knot displays a 
velocity gradient with a smaller blue-shift for larger distance from the star. 

The intensity of the knot is too low to perform a line-by-line measurement of the velocities over the 
feature. Therefore, we added up the ten lines in the spatial direction. The 
resulting nebula plus knot c emission spectrum is shown in Figure \ref{fig:jetspec}. The jet emission 
is very faint and seen only in H$\alpha$ and [N\,{\sc ii}]\,6583 (labeled as J\,H$\alpha$ and 
J\,[N\,{\sc ii}]\,6583 in the plot), while \citet{2016A&A...585A..81M} discovered it also in 
[N\,{\sc ii}]\,6548 and the two [S\,{\sc ii}] lines in their long-exposure image. The strongest jet 
emission line is J\,[N\,{\sc ii}]\,6583 from which we derive a mean knot radial velocity of 
-356.5\,km\,s$^{-1}$,  in agreement with the estimates of \citet{2016A&A...585A..81M} from their 
low-resolution spectra\footnote{Note that \citet{2016A&A...585A..81M} did not correct their 
measurements for the systemic velocity.}.

A blow-up of the jet emission feature J\,[N\,{\sc ii}]\,6583 is shown in the inset of 
Figure \ref{fig:jetspec}. It demonstrates that the line profile of knot c is rather broad and 
double-peaked compared to the narrow, single-peaked regular nebula line. Considering the double-peaked 
profile shape in combination with the elongated inclined structure identified on the 2D image, we tend
to believe that knot c might consist of at least two individual sub-structures. Fitting both with 
individual Gaussians, we obtain a radial velocity of $-378.9$\,km\,s$^{-1}$ and FWHM of 
25.4\,km\,s$^{-1}$ for the narrow blue component, and a radial velocity of $-334.0$\,km\,s$^{-1}$ and 
FWHM of 55.9\,km\,s$^{-1}$ for the broad red component.

We note that no indication for either a blue- or a red-shifted knot component could be identified in 
the FEROS spectra covering the innermost 2\arcsec.

\subsection{Small-scale structures in the near infrared}

\begin{figure}
\includegraphics[width=\hsize,angle=0]{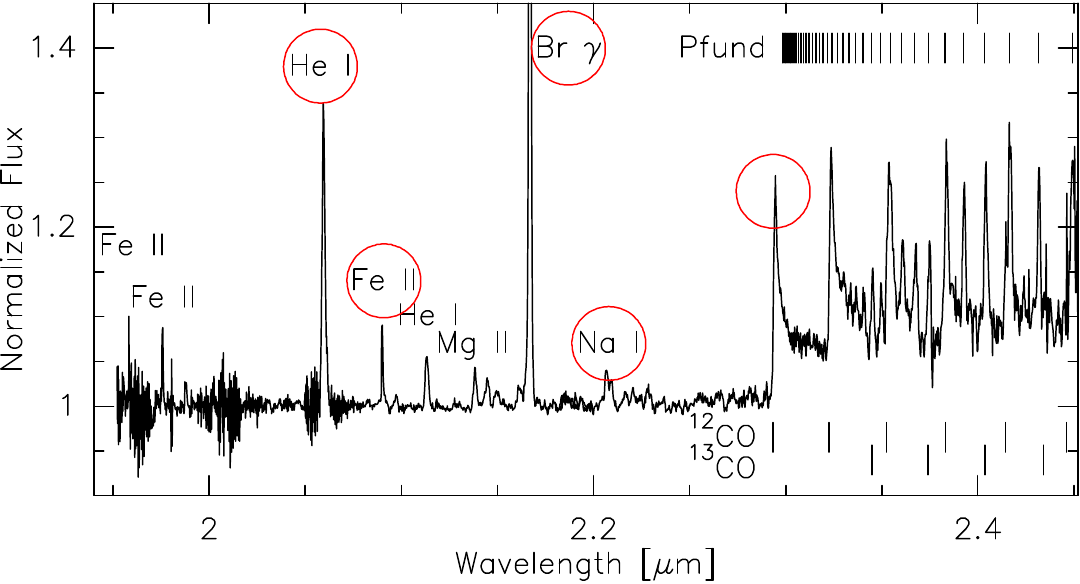}
\caption{SINFONI K-band spectrum. Prominent emission features are marked. The blue region contains 
significant remnants from telluric pollution that could not be cleaned.}
\label{fig:K-spec}
\end{figure}

We now turn to the near-IR spectra that trace the circumstellar material in the close vicinity of
MWC\,137. The full extracted K-band spectrum obtained with SINFONI is shown in Figure \ref{fig:K-spec}. 
It displays numerous emission features and prominent CO band emission. This molecular emission has 
first been detected by \citet{2013A&A...558A..17O}, who observed MWC\,137 with SINFONI using the 
largest spatial plate scale (8\arcsec $\times$ 8\arcsec). The global spectral appearance in both 
SINFONI observations is practically identical. Based on high-resolution observations focused on the CO 
first-overtone bands, \citet{2015AJ....149...13M} resolved the profile of the band heads. Their 
intensities were found to display a blue shoulder and a red maximum, typical for rotating media. With 
the interpretation of the CO bands originating from a Keplerian disk around MWC\,137, 
\citet{2015AJ....149...13M} derived a rotation velocity, projected to the line of sight, of $v\sin i = 
84\pm 2$\,km\,s$^{-1}$. This velocity is too small to be resolved with SINFONI.
 
\begin{figure*}
\begin{center}
\includegraphics[width=\hsize,angle=0]{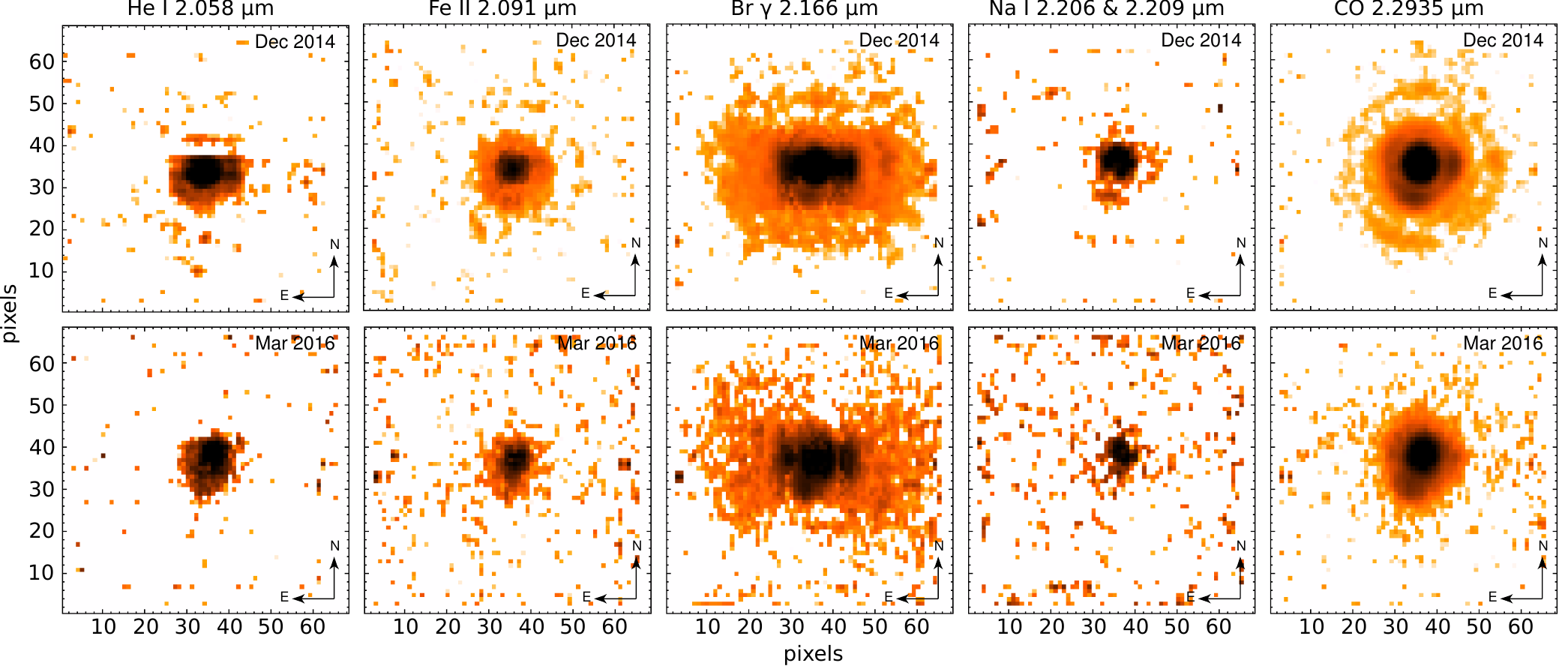}
\caption{Spatial distribution of continuum subtracted line emission of specific features. 
The top row shows the observations from 2014, the bottom row those from 2016 for comparison.
The image sizes are 0\farcs 8 $\times$ 0\farcs 8. To visualize the features, the display has been 
optimized individually for each panel.}
\label{fig:Sinfoni} 
\end{center}
\end{figure*}

With the IFU data cubes in highest spatial resolution, we aimed at tracing the spatial distribution
of the circumstellar gas, and in particular of the hot molecular emission. Therefore, we investigated 
in more detail those individual emission features, which are marked by red circles in the K-band 
spectrum in Figure \ref{fig:K-spec}. For each line, we combined the individual slices from the data 
cube and subtracted a corresponding continuum image. The sodium lines are blended so that we combined 
the emission from both lines into one image. As the red portion of the CO band emission is contaminated 
by Pfund line emission, we used only the non-polluted region around the first band-head. The images of 
the continuum subtracted emission observed in December 2014 and in March 2016 are displayed for 
comparison in the top and bottom row of Figure \ref{fig:Sinfoni}, respectively. 

Starting with the data from 2014, we note that extended emission is seen in all the lines, although the 
emission in the sodium lines appears to be weakest and concentrated around the stellar position. 
Br\,$\gamma$ and CO have clearly structured emission patterns. The Br\,$\gamma$ emission appears to be 
spread over a large volume, but mainly concentrated along the east-west direction with two prominent 
clumps east and west of the star and one towards the south-east. Additional minor, possibly arc-like 
structures appear north and south of these blobs, and a big circular shell-like structure with radius 
$\sim$ 0\farcs 24 might exist as well. A shell (or ring) of similar size is also visible in the CO 
image. Moreover, the CO emission displays three pronounced clumps seen east (least prominent), west, 
and south-east of the stellar position which coincide pretty well with the structures seen in 
Br\,$\gamma$. In fact, the south-eastern clump seems to be present (though weak) in the images of the 
other lines as well. The projected radial distances of these clumps range from 0\farcs 092 to 
0\farcs 118.

The data from 2016 have much worse quality. Still we see the extended emission in both Br\,$\gamma$
and CO. There is one remarkable difference to the images from 2014: The clump to the south-east
seems to have turned eastward by about 10\degr \ in a time span of just 15 months. This clump is again 
visible in the images of all investigated lines.

While the spectral resolution of SINFONI is too low to resolve the kinematical broadening of the CO 
bands, one would still expect to see differences in the spectra extracted at individual pixels due to 
a temperature decrease of the CO gas from inside out.
As the shape of the CO band spectrum, and in particular the relative strength of the individual band 
heads, is sensitive to the gas temperature \citep[see, e.g.,][]{2009A&A...494..253K}, the CO bands
should look considerably different when extracted close to the star compared to those extracted at far 
distances. However, we find that the shapes of the CO band spectra are all identical, and also the 
normalized intensities are the same, independent from the position in the environment from which they 
were extracted. Such a behavior disagrees with the interpretation that the images display CO band 
emission originating from the environment of MWC\,137. Instead, the emission we see appears to be light 
that was reflected by dust, and the clumpy structures we see are presumably dust concentrations 
around the central object. Support for such an interpretation comes from the images of the other
emission features, which all display the same south-eastern clump, independent of the ionization
and excitation state of the element. The image of He{\sc i}, which is also one of the strongest 
features in the spectrum, additionally displays the blobs east and west of the star. Br\,$\gamma$ is 
the most intense spectral feature. Whether its image might be composite, i.e., consisting of reflected 
light and real emission from the ionized wind, or whether it is also just pure reflected light, is 
difficult to judge.

\subsection{Other hot molecules}
 
\begin{figure}
\begin{center}
\includegraphics[width=\hsize,angle=0]{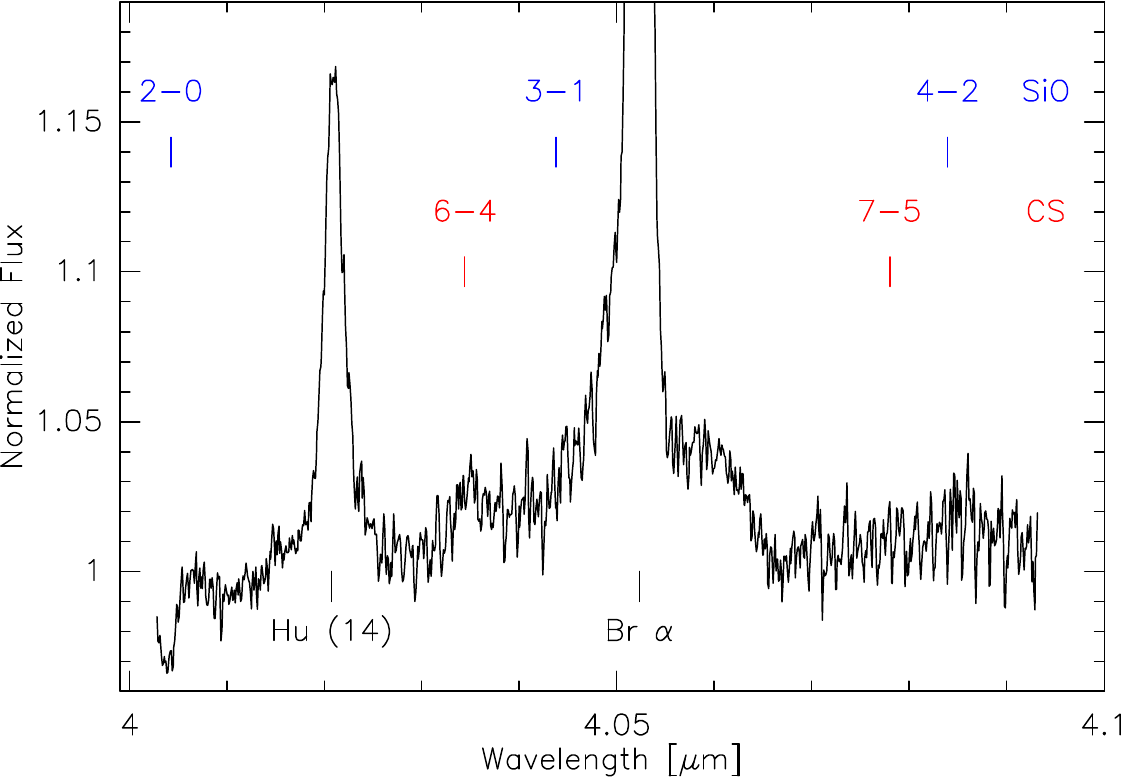}
\caption{High-resolution GNIRS L-band spectrum with two prominent hydrogen recombination lines from the Brackett and Humphreys series. The bottom panel is a zoom to the continuum region. Wavelengths of CS and SiO bandhead positions are marked but not clearly detected.}
\label{fig:N-band} 
\end{center}
\end{figure}

The infrared spectral range is most ideal to search for emission from other molecules. In a pioneering 
study, \citet{2015ApJ...800L..20K} identified SiO band emission from a sample of four B[e] supergiants 
with confirmed CO band emission. These SiO bands arise in the 4\,$\mu$m region around the Br\,$\alpha$ 
line. Our GNIRS spectrum is centered on the Br\,$\alpha$ position, which obviously dominates the 
spectrum (top panel of Figure \ref{fig:N-band}). Also prominent is the emission from the hydrogen line 
Hu(14). To search for molecular features, we zoomed in to the continuum (bottom panel of Figure 
\ref{fig:N-band}) and marked the band head positions of two molecules: SiO and CS. Unfortunately, this
spectral region is strongly polluted by telluric features, which could not be satisfactorily removed, 
hampering significantly the proper identification of molecular band head structures. We 
computed a large grid of model spectra for both molecules but could not find any reasonable agreement 
with the features. We conclude that if present, the emission features from any of these molecules must
be negligibly weak.

%-------------------------------Fig. 
\begin{figure}
\begin{center}
\includegraphics[width=\hsize,angle=0]{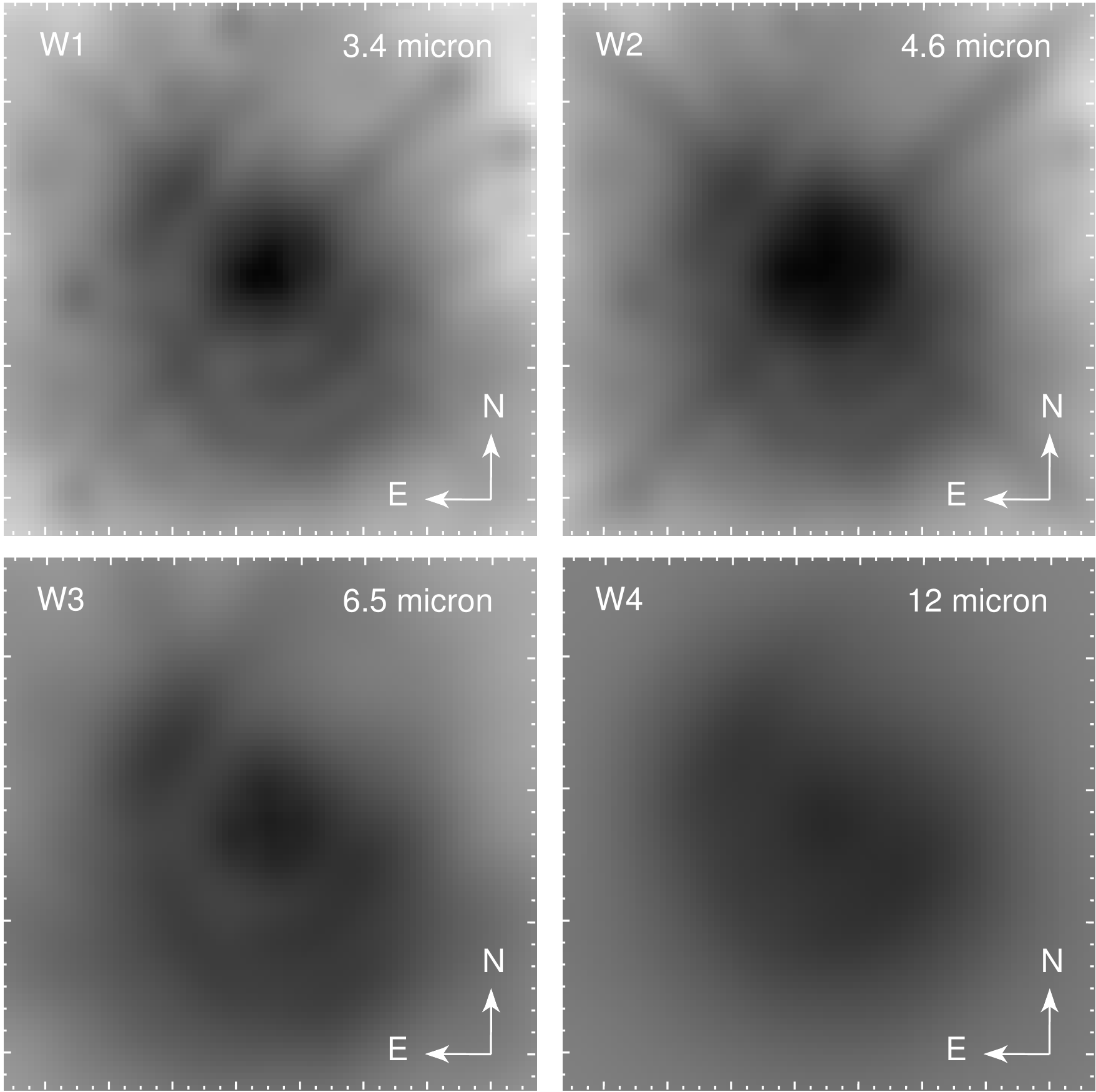}
\caption{WISE images of MWC\,137 in the four different bands. Image sizes are 
2\arcmin $\times$ 2\arcmin.}
\label{fig:WISE} 
\end{center}
\end{figure}
%------------------------------------- 

\subsection{Circumstellar dust}\label{Sect:dust}

The presence of dust around MWC\,137 is known from the recorded large infrared excess emission 
\citep{1972ApL....11...95F, 1975ApL....16..165C, 1992ApJ...397..613H}. However, not much is 
known about the spatial distribution of the dust. This can be investigated based on the numerous 
infrared images that were acquired in public surveys during the past few years.

MWC\,137 was observed in all four WISE bands. The images are shown in Figure \ref{fig:WISE}. 
Arc-like structures around the central star are resolved in all four bands. As the angular resolution 
of WISE drops significantly with increasing wavelength, this structure appears rather blurry at 
12\,$\mu$m. Extended emission of similar size and location is also seen in the far-infrared 
{\em Herschel} images acquired with PACS and SPIRE. The detectability of intense, extended emission up 
to 500\,$\mu$m means that the ionized nebula of MWC\,137 is surrounded by significant amounts of warm 
and cool dust. With the poor angular resolution of the far-infrared images, the emission appears blurry
similar (or worse) to the W4 image shown in Figure \ref{fig:WISE} and we refrain from showing them 
here.

It is noteworthy that the two images in W1 and W2 look very much alike. This finding is
confirmed by the equal appearance of the two GLIMPSE survey images taken with much better angular 
resolution and sensitivity in the passbands IRAC-1 and IRAC-2 centered at 3.6\,$\mu$m and 4.5\,$\mu$m. 
Both bands cover specific but distinct emission features \citep[e.g.,][]{2008ApJ...681.1341W}:
IRAC-1/W1 covers the emission of PAHs at 3.3\,$\mu$m, while IRAC-2/W2 covers emission from shocked 
molecular gas, such as H$_{2}$ lines and CO ro-vibrational transitions of the fundamental band as is 
found for example in the outflows of massive young stellar objects \citep{2008AJ....136.2391C} and/or 
emission of Br$\gamma$ and Pf$\beta$ from ionized gas. Therefore, if either PAH emission or emission 
from shocked molecular gas/ionized gas would be present in the nebula of MWC\,137 and dominate the 
near-infrared appearance, the spatial distribution of the emission in these two bands should be 
physically disjoint. As this is not the case, we conclude that the near-infrared 
%images are neither dominated by PAHs nor by emission from shocked 
%molecular gas. Instead, the infrared 
emission structures most likely trace the spatial distribution of purely thermally emitting dust.
The absence of shocked molecular gas is supplemented by the non-detection of H$_2$ emission lines in 
the K-band spectrum (see Figure \ref{fig:K-spec}).
 
%-------------------------------Fig. 
\begin{figure}
\begin{center}
\includegraphics[width=\hsize,angle=0]{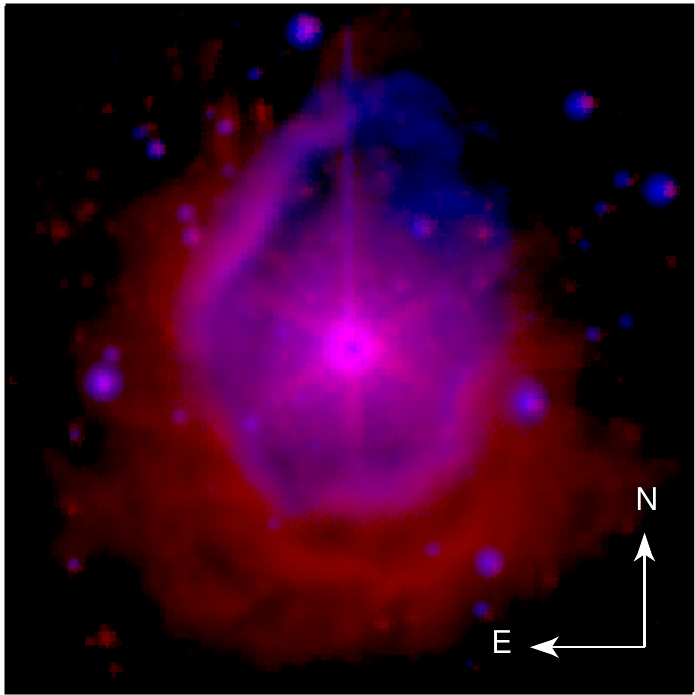}
\caption{Composite image of the ALFOSC H$\alpha$ (blue) and the Spitzer 3.6\,$\mu$m (red) images. 
Image size is 2\arcmin $\times$ 2\arcmin.}
\label{fig:Spitzer} 
\end{center}
\end{figure}
%-------------------------------------

Using the high-quality near-infrared images we can compare the spatial extend of both 
dust and ionized gas. Figure \ref{fig:Spitzer} shows a composite image of GLIMPSE 3.6\,$\mu$m
(red) and ALFOSC H$\alpha$ (blue). Comparison of the near-infrared emission to the optical one reveals, 
that the dust emission is basically encircling the optical nebula, following the wiggly structure 
of the optical filaments. However, we note two significant differences. There is no dust emission 
towards the north-west direction. Moreover, in the south the dust emission displays an additional 
arc-like structure which has no optical counterpart. This additional dust arc appears at a distance of 
about 40\arcsec and is hence 10\arcsec -- 15\arcsec\ (from south to south-west) farther out than the 
optical southern filament.

\subsection{Cold Molecular gas}

We turn now to the description of the radio data. These contain the information of the cold molecular
gas at far distances from the star. 

\begin{figure}
   \centering
   \includegraphics[width=\hsize,angle=0]{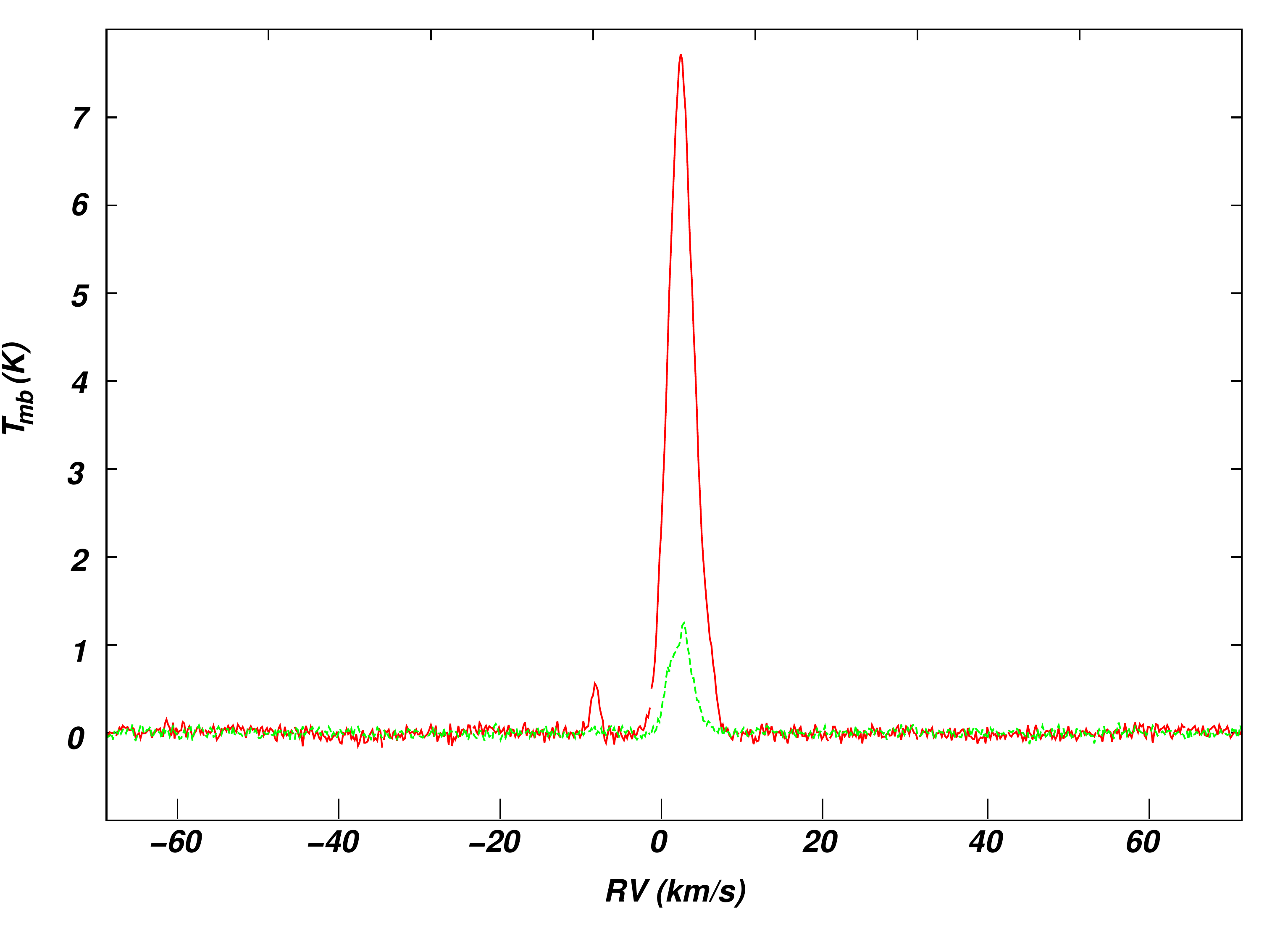}
   \caption{Averaged $^{12}$CO(3-2) (red) and $^{13}$CO(3-2) (green) spectra in units of the main-beam 
   brightness-temperature, $T_{\rm mb}$,  obtained within the observed area}
   \label{spectrum}%
    \end{figure}

The $^{12}$CO(3-2) and  $^{13}$CO(3-2) spectra averaged within a 3\arcmin$\times$3\arcmin\ region 
centered on RA(J2000) = 6$^h$18$^m$45.5$^s$ and Dec(J2000) = +15\degr16\arcmin52\farcs 3 are shown in 
Figure \ref{spectrum}. The molecular emission accumulates in one bright component between $-$2 and 
+8\,km\,s$^{-1}$ as detected in the $^{12}$CO(3-2) data, and between --1.0 and +5.8\,km\,s$^{-1}$ in  
$^{13}$CO(3-2) emission. A very faint component detected at --8\,km\,s$^{-1}$ in $^{12}$CO(3-2) has no 
clear relation to the nebula and will be ignored in the present analysis. No obvious signal was 
detected in either the SiO(5-4) or CS(5-4) molecular lines.

\subsubsection{Spatial distribution}

\begin{figure*}
   \centering
\includegraphics[width=0.95\hsize,angle=0]{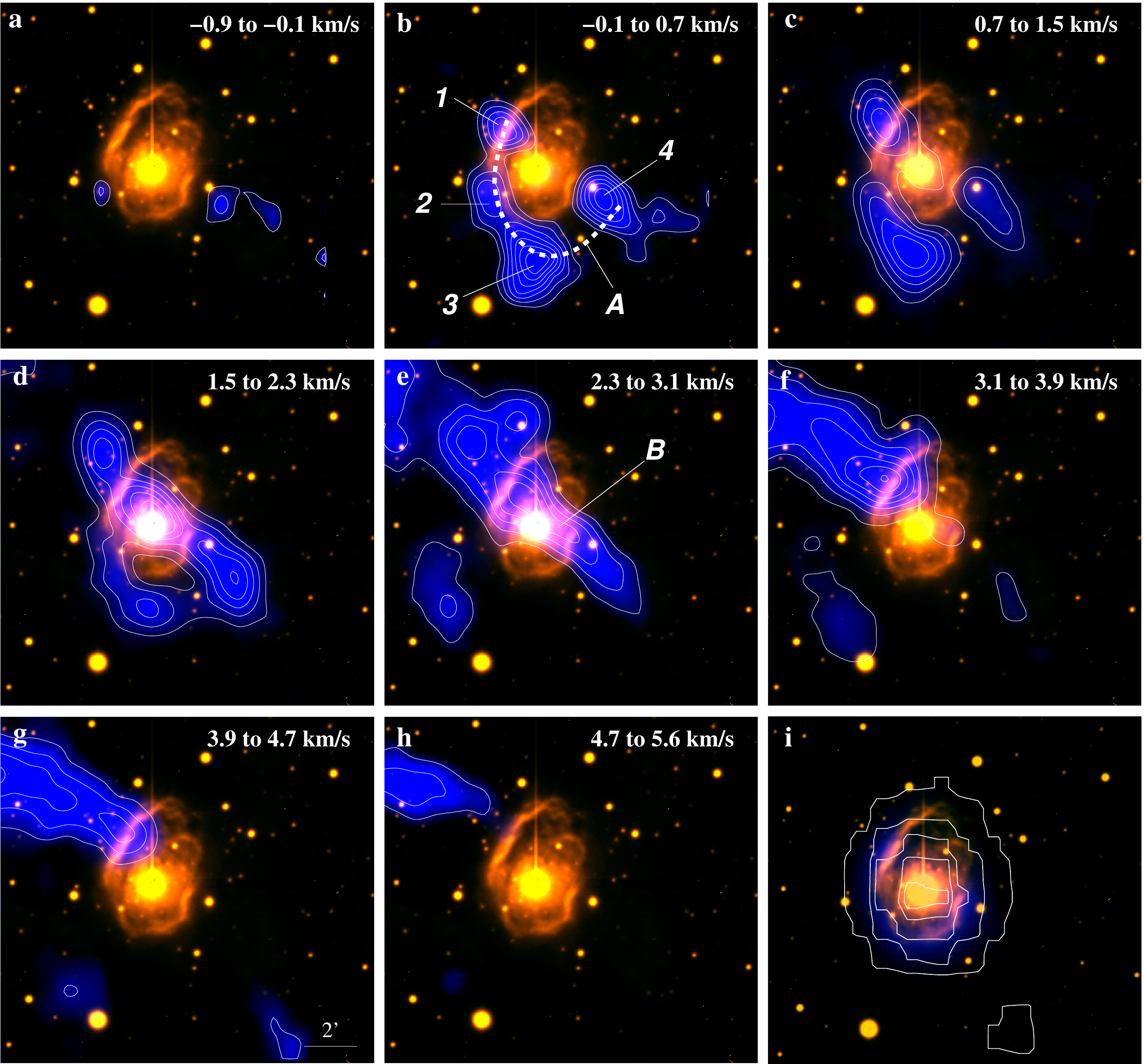}
   \caption{$^{13}$CO(3-2) line emission distribution (in blue) in the velocity range from --0.9 to 
  +5.6\,km\,s$^{-1}$ in steps of 0.8\,km\,s$^{-1}$, superimposed onto the ALFOSC H$\alpha$ image (in 
  orange). The numbers in panel {\it b} mark the position of the four clumps in Structure A. Velocity 
  interval of each panel: a) [--0.9,--0.1]\,km\,s$^{-1}$; b) [--0.1,+0.7]\,km\,s$^{-1}$; 
  c) [+0.7,+1.5]\,km\,s$^{-1}$; d) [+1.5,+2.3]\,km\,s$^{-1}$; e) [+2.3,+3.1]\,km\,s$^{-1}$; 
  f) [+3.1,+3.9]\,km\,s$^{-1}$; g) [+3.9,+4.7]\,km\,s$^{-1}$; h) [+4.7,+5.6]\,km\,s$^{-1}$. 
  Contour levels for {\it a}  and {\it b}: 0.6 (3 rms) to 2.1 K in steps of 0.3 K; for {\it c}  to 
  {\it h}: from 1.5 to 8.5 K in steps of 1.0 K. The last panel i) displays an overlay of the 
  Herschel-PACS emission at 70\,$\mu$m (in blue) and the NVSS 1420\,MHz radio continuum data (white contours at 
  0.01, 0.02, 0.03, 0.04, and 0.047\,Jy/Beam) on the  H$\alpha$ image. 
  }
   \label{mosaico}%
    \end{figure*}

The structures seen in the two CO isotopes are generally the same, with the $^{12}$CO(3-2) emission
slightly more extended than the $^{13}$CO(3-2) emission due to the presence of low-intensity diffuse 
gas. But since the $^{12}$CO(3-2) emission is optically thick, it provides limited information on the 
internal structure and dynamics of the molecular gas. Therefore, we focus on $^{13}$CO(3-2) and 
integrate the emission in the velocity interval [$-$0.9,+5.5]\,km\,s$^{-1}$, in steps of 
0.8\,km\,s$^{-1}$. The distribution of the emission within these velocity bins is displayed in Figure 
\ref{mosaico}. The emission concentrates at and around the eastern, southern and south-western borders 
of the optical nebula and is in perfect alignment with the spatial distribution of the dust identified 
in the near- and mid-IR (see Figure \ref{fig:WISE}). Alike for dust, there is an obvious lack of 
molecular gas towards the northwest, coincident with the region of largest negative nebular velocities 
at PA = 35 (see Figure \ref{fig:rv}).

We detect two extended structures towards the nebula. The first one (henceforth Structure A)
is present in the velocity range [$-$0.1,+1.9]\,km\,s$^{-1}$ (panels {\it b} to {\it d}) and 
encircles the eastern and southern optical rims. This structure is subdivided into four molecular 
clumps as obvious from panel {\it b}. Comparison with the optical images shows that clump 1 
coincides with a bright optical filament, while clumps 2 and 3 
have their emission peaks to the south-east and south of, but outside the optical nebula. These 
latter two clumps (2 and 3) merge into an arc-like structure at higher velocities (see panels {\it c} 
and {\it d}). The center of clump 3 coincides with the additional arc-like structure mentioned in the 
last paragraph  of Sect.\,\ref{Sect:dust}. A fourth clump (4) 
closely borders the brightest optical filament to the west. The spatial correlation of 
this partial ring-like structure and the H$\alpha$ emission strongly suggests that material at these 
velocities corresponds to the molecular counterpart of the optical nebula.

%-------------------------------Fig. 
%-------------------------------------
%------------------------------------Fig. 
%-----------------------------------------------

The second configuration (Structure B) is an elongated feature that runs across the nebula in the  
northeast-southwest direction (PA $\simeq$ 310; panels {\it d} to {\it f}), approximately perpendicular 
to the major axis of the elongated optical nebula. The entire structure is detected from 
+3.9\,km\,s$^{-1}$ down to +1.5\,km\,s$^{-1}$ where it breaks apart and ends up in clumps 1 and 4. 
The regions of maximum intensity in Structure B coincide with the brightest domains in the filaments of 
the optical nebula Sh\,2-266. 
The north-eastern part of Structure B is still visible in the velocity interval 
[+4.1,+5.5]\,km\,s$^{-1}$ (panels {\it g} and {\it h}), which is just a bit higher. Considering
the huge spatial extend of this whole elongated feature which displays only a small variation
in velocities, we might conclude that Structure B encloses a rather small angle with the plane
of the sky.
 
A comparison of the  70\,$\mu$m emission detected by Herschel-PACS is displayed in 
panel {\it i} of  Figure \ref{mosaico}. It demonstrates that the far IR emission of the cold dust 
coincides\footnote{We note that the emission at 160\,/$\mu$m (not shown) is slightly more extended.} 
with the molecular emission, providing a link between dust and molecular gas in the two Structures A 
and B.

%--------------------------------Table 
\begin{table*}
\caption{Parameters of the cold molecular gas\label{tab:medidas}}
\centering
\begin{tabular}{lcccccccccc}
\hline
Structure & $T^p_{\rm 12CO}$ & $T^p_{\rm 13CO}$ & $T_{\rm exc}$ & $\tau_{13}$  & $T_{\rm mean}$ & $\Delta {\rm v}$ & $N_{\rm 13CO}$ & $N_{\rm H2}$ & $R_1\times R_2$ & $M_{\rm H2}$  \\
   & (K) & (K) & (K) &   &  (K) & (km\,s$^{-1}$) & (10$^{15}$cm$^{-2}$)& (10$^{21}$cm$^{-2}$)& (\arcsec) & (M$_\odot$)    \\   
\hline 
\hline
A-1         & 18.4 & 4.3 & 25.8 & 0.26 & 1.11 & 2.4 & 1.46 & 1.1 & 31$\times$21 & 35 \\
A-2-3       & 14.5 & 2.8 & 21.7 & 0.21 & 1.64 & 2.2 & 2.03 & 1.6 & 48$\times$25 & 85  \\
A-4         & 14.0 & 3.1 & 21.2 & 0.25 & 0.94 & 1.8 & 0.98 & 0.8 & 21 & 15    \\
B-southern  & 18.0 & 3.3 & 25.4 & 0.20 & 1.09 & 4.4 & 3.37 & 2.6 & 40$\times$25 & 120  \\
B-northern  & 15.0 & 3.5 & 22.2 & 0.26 & 2.16 & 4.4 & 5.47 & 4.2 & 39$\times$34 & 245  \\
\hline
\end{tabular} 
\label{medidas}
\end{table*}
%----------------------------------------------

\subsubsection{Physical parameters of the molecular components}

We evaluate the physical parameters of the two identified structures separately, although they are 
partially superposed in velocity and in space. For the arc-like Structure A that surrounds Sh\,2-266 
we consider the  $^{13}$CO emission separately for clump 1, clumps 2 plus 3, and clump 4. For the 
elongated Structure B we individually derive the values for the northern and southern sections. 

Assuming Local Thermodynamic Equilibrium (LTE) conditions and that the emission in the $^{12}$CO line 
is optically thick, we compute the  excitation temperature T$_{\rm exc}$ from the emission in the 
$^{12}$CO(3-2) line. Using the relation
\begin{equation}
\label{eq:texc}
  T^p_{\rm 12CO} \ [{\rm K}]\ =\ T_{\rm 12CO}^*\ [J_{12}(T_{\rm exc}) \ -\ J_{12}(T_{\rm bg})]
\end{equation}
for the  peak main-beam brightness-temperature ($T^p_{\rm 12CO}$) of the $^{12}$CO line, where 
$T_{\rm 12CO}^*$ = $h \nu_{12}/ k$ with the frequency $\nu_{\rm 12}$  of the $^{12}$CO(3-2) line,  $J_{12}(T) = (e^{T_{\rm 12CO}^* / T} -1)^{-1}$, and  $T_{\rm bg}$ is the background 
temperature, for which we use the value of 2.7 K, the excitation temperature for this CO line results to 
\begin{equation}
\label{eq:texc2}
T_{\rm exc} \ [{\rm K}]\ =\ \frac{16.59}{{\rm ln}\ [ 1\ +\ 16.59\ / \ (T^p_{\rm 12CO}\ + 0.036) ] }\,.
\end{equation}
The values for $T^p_{\rm 12CO}$ are thereby individually measured for the different clumps
and structures from the averaged spectra.

The optical depth $\tau_{13}$ was obtained from the \hbox{$^{13}$CO(3-2)} line  using the expression 
\begin{equation}
\tau_{13} = -\ln\left[1-\frac{ T^p_{\rm 13CO}}{T_{\rm 13CO}^*}
\left[J_{13}(T_{\rm exc})-J_{13}(T_{\rm bg})\right]^{-1}\right]\,,
\label{tau13co}
\end{equation}
where $T_{\rm 13CO}^*$ = $h \nu_{\rm 13} / k$ with the frequency $\nu_{\rm 13}$  of the $^{13}$CO(3-2) 
line, and $J_{13}(T) = (e^{T_{\rm 13CO}^* / T} -1)^{-1}$. For this derivation, we assumed that the 
excitation temperature for the $^{13}$CO emission line is the same as for the $^{12}$CO line. The 
$^{13}$CO line $(T^p_{\rm 13CO})$ peak main-beam brightness-temperatures for the clumps in Structure A 
and for the northern and southern sections of Structure B were also obtained from the averaged spectra 
within the emitting regions. Peak and excitation temperatures, and optical depths are summarized in 
cols.~2, 3, 4, and 5 of Table\,\ref{tab:medidas}.

Assuming LTE, the $^{13}$CO column density, $N_{\rm 13CO}$ [cm$^{-2}$], 
can be estimated from the $^{13}$CO(3-2) line, following Buckle et al. (2010),
\begin{equation}
N_{\rm 13CO} = 8.3\times 10^{13} e^{15.87/T_{\rm exc}} \frac{T_{\rm exc} + 0.88}{1 - e^{-15.87/T_{\rm exc}}}  \int \tau_{13} d{\rm v}\,. 
\label{n13co}
\end{equation}
The integral in Eq. (\ref{n13co}) can be approximated by
\begin{equation}
\int\tau_{13} d{\rm v} = \frac{1}{T_{\rm 13CO}^* [J(T_{\rm exc}) - J(T_{\rm bg})]} 
\frac{\tau_{13}}{1 - e^{-\tau_{13}}}  \int T_{\rm mb}\  d{\rm v}\, ,
\label{integral}
\end{equation}
with
\begin{equation}
 \int{T_{\rm mb}}\  d{\rm v}  = T_{\rm mean} \Delta {\rm v}.
\label{integral2}
\end{equation}
Eq. (\ref{integral}) is appropriate  to eliminate small optical depth effects and is good within 15\% 
for the derived values of $\tau_{13}$ \citep{2004tra..book.....R}.  Bearing in mind the 
$\tau_{\rm 13}$-values listed in Table  \ref{tab:medidas}, the correction increases the optically thin 
column density by about 10-15\%. $T_{\rm mb}$ and $T_{\rm mean}$ in Eq. (\ref{integral2}) are the 
main-beam and the averaged brightness temperature  of the $^{13}$CO(3-2) line, and $\Delta {\rm v}$ is
the velocity interval of the emitting areas.
For the arc-like Structure A that surrounds Sh\,2-266 
we consider the  $^{13}$CO emission within the velocity interval 
\hbox{[$-$0.9,+1.5]\,km\,s$^{-1}$} (panels {\it a} to {\it c} in Figure \ref{mosaico}) for clump 1, 
[$-$0.3,+2.1]\,km\,s$^{-1}$ (panels {\it a} to {\it d}) for clumps 2 and 3, and 
\hbox{[$-$0.9,+0.9]}\,km\,s$^{-1}$ (panels {\it a} to {\it c}) for clump 4. 
For the elongated Structure B we take into account the emission in the velocity range 
[+0.9,+5.7]\,km\,s$^{-1}$ and  derive the values for the northern and southern sections. 
   
The molecular mass was calculated using
\begin{equation}\label{eq:masa}
M_{\rm H2}\ [{\rm g}] \ =\ \mu\ m_H\  A \ N_{\rm H2}\ d^{2}
\end{equation}
where $\mu$ is the mean molecular weight, which is assumed to be equal to 2.72 after allowance of a 
relative helium abundance of 25\% by mass \citep{1999sf99.proc..383Y}, $m_{\rm H}$ is the atomic 
hydrogen mass, 
%($\sim$1.67$\times$10$^{-24}$ g),   
$N_{\rm H2}$ is the H$_{2}$ column density, obtained using an abundance ratio 
\hbox{$N_{\rm H2}$ / $N_{\rm 13CO}$} = 7.7$\times$10$^{5}$ \citep{1994ARA&A..32..191W}. $A$ is the 
solid angle of the CO emitting regions, obtained from the semi-axis $R_1\times R_2$ of the regions, and 
$d$ is the distance to the star. 

The distance to the molecular gas around Sh\,2-266 can be estimated from the velocity field of the 
outer Galaxy, which was derived by \citet{1993A&A...275...67B}. Considering a mean radial velocity of 
$-$0.5\,km\,s$^{-1}$ of Structure A (corresponding to an LSR velocity of +27.5\,km\,s$^{-1}$), we find 
a kinematical distance $d \geq$ 5.5\,kpc. This distance is similar to the spectroscopic distance 
towards the nebula, for which \citet{2016A&A...585A..81M} determined a value of 5.2$\pm$1.4\,kpc, and 
we adopted the spectroscopic value of 5.2\,kpc for our computation of the molecular mass. Uncertainties 
in the molecular masses are about 50\%\,and originate mainly from the distance uncertainties.  

The summary of our results is presented in Table \ref{tab:medidas}, where we list the average main-beam 
brightness-temperature ($T_{\rm mean}$, col.~6), the velocity interval ($\Delta {\rm v}$, col.~7), the 
column densities of $^{13}$CO and H$_2$ (cols.~8 and 9), the semi-axis of the emitting area ($R_1\times 
R_2$, col.~10), and the  molecular mass ($M_{\rm H2}$, col.~11). The huge values found for the 
molecular masses imply that the cold material surrounding MWC\,137 must be interstellar in origin. 

Finally, we compute the volume density $n_{\rm H2}$. This is not an easy task since it involves the 
geometry and volume of the structures. A rough estimate of this parameter for the clumps in Structure A 
adopting the semi-axis listed in Table  \ref{tab:medidas} indicate values in the range 400 to 
660\,cm$^{-3}$. For Structure B we adopt an elongated ellipsoid including the northern and southern 
sections and find a volume density of  1200\,cm$^{-3}$. Uncertainties in these values are about 70\%. 
The relatively low volume densities are compatible with the non detection of high density tracers, 
such as the CS molecule.

\subsection{Ionized gas}

Alike the H$\alpha$ image the radio continuum emission at 1.4 GHz indicates the presence of 
ionized gas (Fig.~\ref{mosaico}). The image acquired during the NVSS survey reveals a centrally 
peaked source coinciding with Sh\,2-266, with a flux density $S_{1.4\,\rm GHz}$ =  78 mJy.

The number of Lyman continuum photons needed to maintain the ionization of the H{\sc ii} region can be 
calculated from the 1.4 GHz emission  using
\begin{eqnarray}
\quad    N_{\rm Lyc}\ [{\rm s}^{-1}]=\ 7.58\ \times\ 10^{48}\      T^{-0.5}_{\rm e}\  S_{1.4\,\rm GHz}\ d^2
\label{nlym}
\end{eqnarray}
\citep{1994ApJS...91..659K}. Adopting an electron temperature of 10$^4$\,K, we estimate $N_{\rm Lyc}$ = 
1.6$\times$10$^{47}$  s$^{-1}$. Assuming that about 50$\%$ of the UV photons are absorbed by 
interstellar dust in the H{\sc ii} region \citep{2001AJ....122.1788I}, a Lyman continuum flux of 
3.2$\times$10$^{47}$ s$^{-1}$ is necessary. This value should be easily provided by a B0.5I star, with 
a UV photon flux of 4$\times$ 10$^{47}$\,s$^{-1}$ \citep{2002MNRAS.337.1309S}. Bearing in mind the
presence of a stellar cluster in the center of the nebula \citep{2016A&A...585A..81M}, this value could
be a lower limit.

\section{Discussion}

\begin{figure}
\begin{center}
\includegraphics[width=\hsize,angle=0]{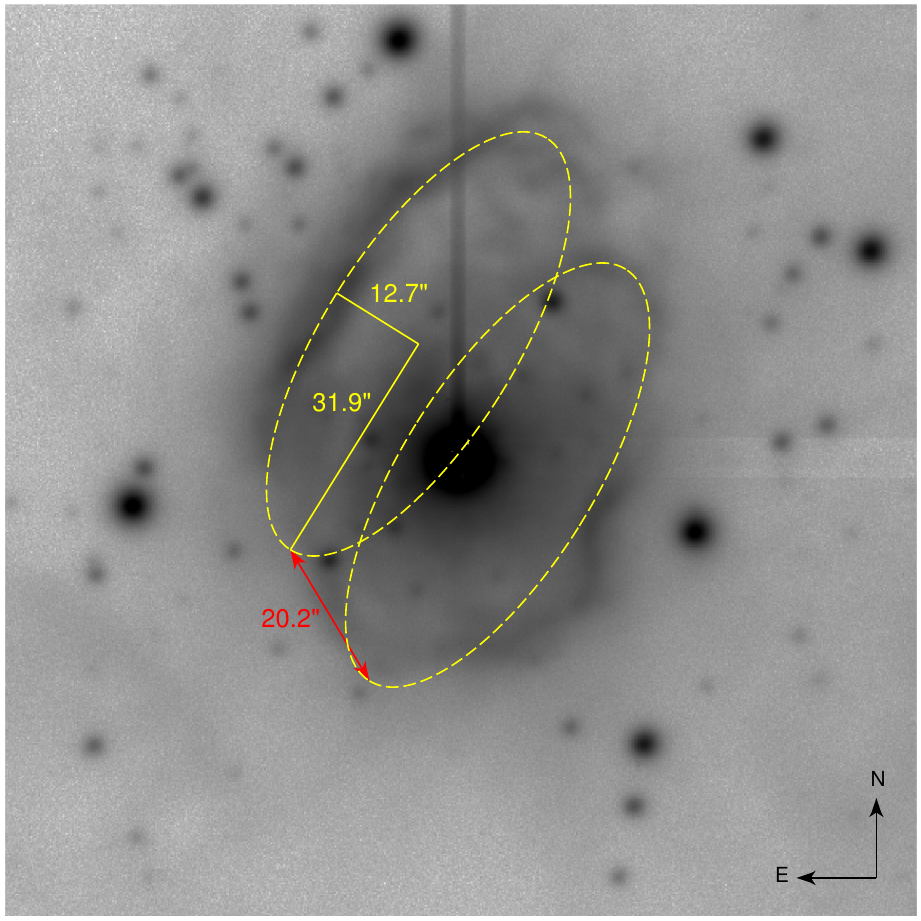}
\caption{Measured quantities of the putative double-ring (or cone) structure of the optical nebula.
The image size is 2\arcsec $\times$ 2\arcsec.}
\label{fig:double_ring} 
\end{center}
\end{figure}

The optical nebula displays some structure that might be approximated by a double-ring. This is 
visualized by the two identical, parallel ellipses overlaid on the ALFOSC image in Figure 
\ref{fig:double_ring}. We measured the semi-major ($a$=31\farcs9$\pm$1\farcs0) and semi-minor 
($b$=12\farcs7$\pm$0\farcs5) axis, the alignment of the ellipses (here: of the semi-minor axis 
PA($b$)=328\fdg90$\pm$0\fdg50), the distance ($c$=20\farcs2$\pm$0\farcs2) between the two ellipses, 
and the alignment of the line connecting the two ellipses (PA($c$)=301\fdg89$\pm$0\fdg50). 

The equal dimensions of the ellipses and their parallel arrangement suggest that they might represent
the outer rims of a double-cone. However, since the alignment of the semi-minor axis is not parallel 
to the alignment of the line connecting the two ellipses, this double-cone must be sheared in one
direction. 

To test whether such a structure might be a reasonable scenario to explain the observed nebula 
kinematics, we use the observed parameters as constraints and compute the geometric shape of a sheared 
double-cone. We start by defining a double-cone in north-south direction, which we align with the 
z-axis. The top and bottom circle have radius $R$ which is given by the length of the semi-major axis 
$a$ of the ellipse. The height of each cone is $h$ so that the object extends from $z = -h$ to $z= +h$, 
and the angle $\alpha$ represents half the opening angle of the cone. Then we apply a shear $\Delta y$ 
along the $y$-axis. This shear is characterized by the angle $\xi$ defined via $z = \Delta y \tan\xi$.
Finally, we rotate the double-cone first around the $y$-axis with angle $\theta$ and then around the 
$z$-axis with angle $\phi$, where $\theta$ and $\phi$ represent the usual spherical coordinates. This 
rotated double-cone structure is then projected into the $y-z$ plane, which we identify with the plane 
of the sky.

\begin{figure*}
\begin{center}
\includegraphics[width=0.85\hsize,angle=0]{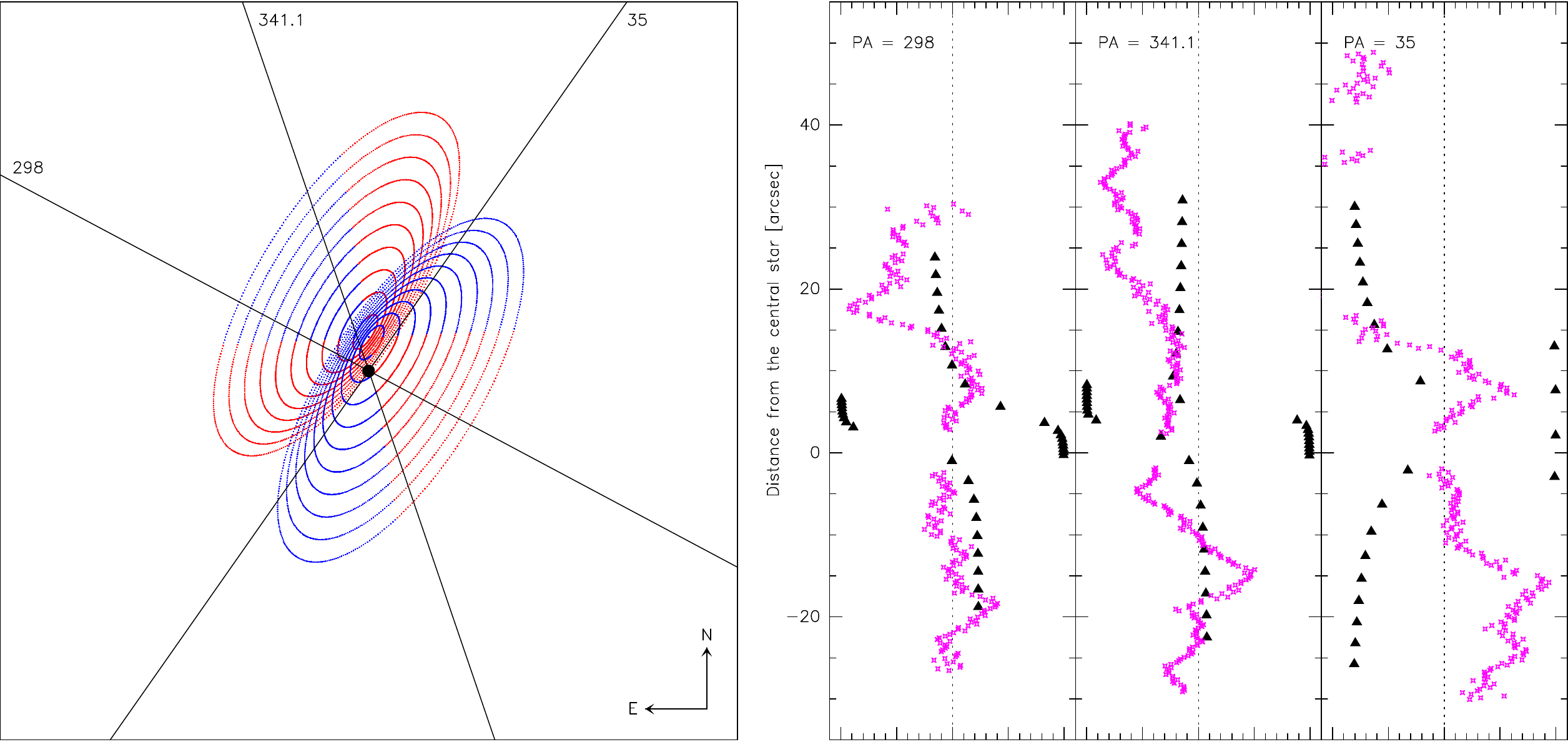}
\includegraphics[width=0.85\hsize,angle=0]{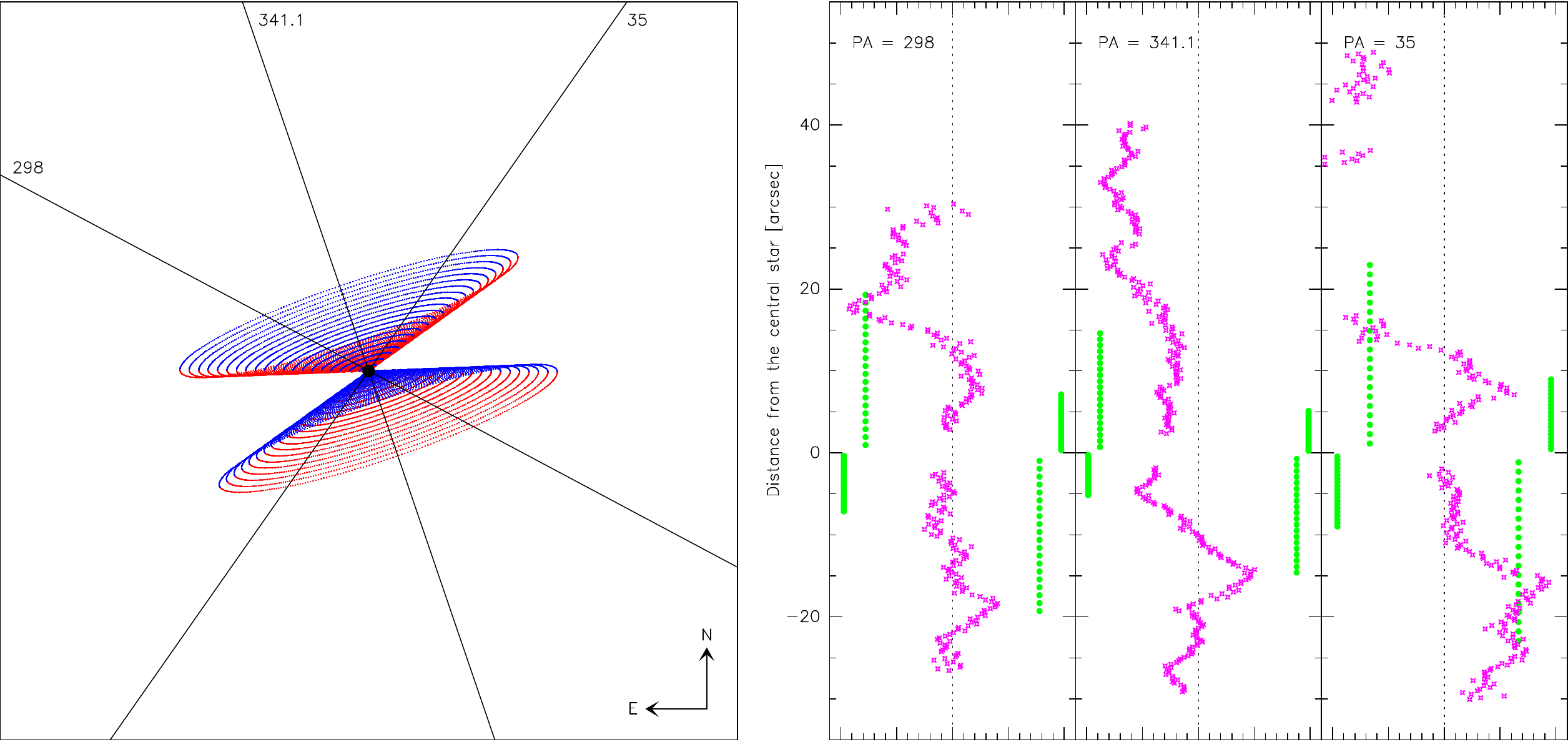}
\includegraphics[width=0.85\hsize,angle=0]{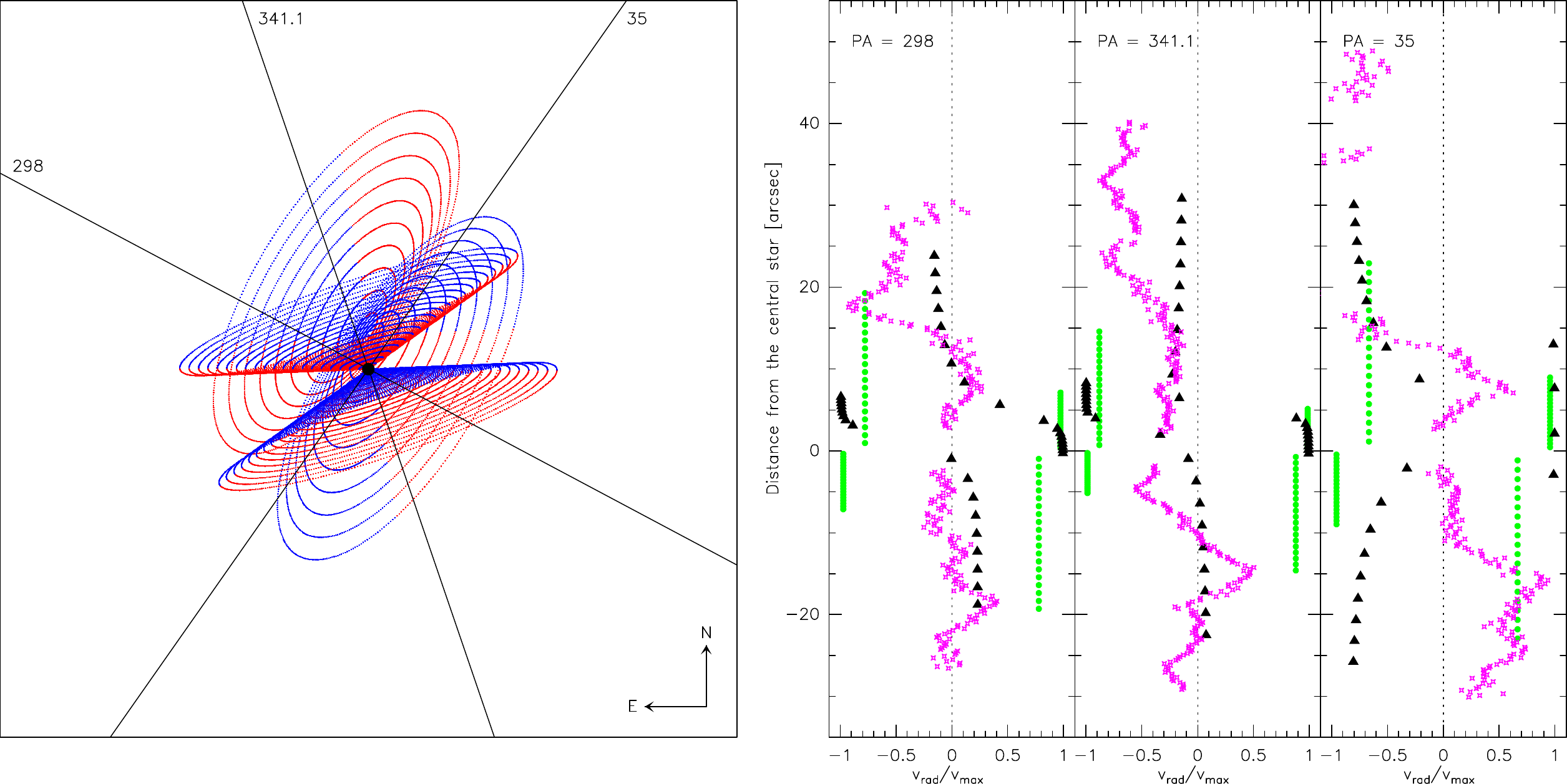}
\caption{{\it Left:} Nebula kinematics for the scenario of a sheared double-cone structure as estimated 
from the optical nebula (top), for a regular double-cone centered on the stellar position and with the 
jet PA as symmetry axis (middle), and a combination of both (bottom). In both cases, the north-eastern 
cone points away and the south-western one towards the observer. Colors indicate regions with 
blue- and red-shifted velocities. Image size corresponds to 2\arcsec $\times$ 2\arcsec. 
{\it Right:} Model kinematics along the PAs (black triangles and green dots) compared to observations 
(purple crosses).}
\label{fig:cone} 
\end{center}
\end{figure*}

To compute the kinematics, we assume that the gas is streaming radially with constant velocity and 
purely along the cone surfaces. This allows us to determine the regions of blue- and red-shifted 
emission. To reproduce the observed quantities and position angles we found the values of
$h$=17\farcs81, $\alpha$=60\fdg86, $\xi$=137\degr \ for the double-cone parameters, and  
$\theta$=119\degr \ and $\phi$=63\degr \ for the rotation angles. The projection of this geometrical
shape to the plane of the sky is shown in the top left panel of Figure \ref{fig:cone}. In blue and 
red we mark regions with blue- and red-shifted gas kinematics, respectively.

Next, we overlaid the position angles of the slits and determined the intersection points with 
the projected double-cone. During the observations the three PAs were positioned such that 
they intersect at the location of the star. However, with respect to the chosen geometrical
model it is important to note that the center of the double-cone was found to be located 5\farcs 5 \
north of the stellar position. The extracted velocity values at the intersection points, normalized to 
the constant outflow velocity, are shown in the top right panel of Figure \ref{fig:cone}. 
For comparison we included the observed velocities, which we normalized arbitrarily.

In the northern nebula regions our model produces blue-shifted emission in agreement with the 
observations. Also the red-shifted northern emission in the vicinity of the star along PA=298 agrees 
with the model predictions. Note that the densely populated regions at distances 0\arcsec--10\arcsec \
north of the star along PA=298 and PA=341.1 with both high blue- and red-shifted velocities originates 
from the overlap regions of the two cones. 
What cannot be reproduced is the variability in the velocities and the high 
amplitudes observed in the blue-shifted gas. In the southern nebula our model produces red-shifted
emission. There is some agreement with the observations along PA=298 and PA=341.1, however,
for PA=35 the observations display strong red-shifted emission in contrast to the theoretical
predictions.

Although some of the kinematics can be reproduced, we are aware that the chosen geometrical scenario is 
too simple to account for all observed features. Still, this exercise provides important insight.
The fact that the star MWC\,137 is not located in the center of the geometrical double-ring structure
suggests that the large-scale optical nebula was most probably not formed exclusively by the
wind of a single star but rather by the combined winds of the early-type stars identified by 
\citet{2016A&A...585A..81M} in the center of the cluster and in the vicinity of MWC\,137. In addition, 
the jet which was found by \citet{2016A&A...585A..81M} to originate from MWC\,137, is not aligned with 
the axis of the double-cone. Instead, the jet is perpendicular to the circumstellar disk of MWC\,137 
which is traced by the intrinsic polarization in H$\alpha$ \citep{1999MNRAS.305..166O} and the 
rotationally broadened CO bands \citep{2015AJ....149...13M}. Under the presence of a circumstellar 
disk, the current wind of MWC\,137 is most likely also channelled into a double-cone, but with a 
different orientation. One might speculate whether this inner disc and the jet might point towards 
MWC\,137 having a compact companion as, e.g., in the case of CI\,Cam 
\citep[see, e.g.,][]{2006ASPC..355..269C}. The detection of an X-ray source in the vicinity of MWC\,137 
by \citet{2014ApJS..210....8E} might support such a suggestion. However, the uncertainty in 
position of the X-ray source renders it difficult to unambiguously identify MWC\,137 as the optical 
counterpart. 

Unfortunately, the optical image provides no clear indication of the postulated second double-cone, 
which might imply a (much) lower wind density. Besides the position angle of the jet and the request 
that the velocity in the southern part of the slit PA=35 should be red-shifted, we have no additional
observational constraints for a possible model. Therefore, and simply for demonstration purposes, we
compute a circular double-cone centered on the star's position. To align it with the PA of the jet we 
only altered the rotation angle $\theta$, for which we find a value of 159\degr \ but kept the same 
value for $\phi$. For the height and the cone opening angle we adopted values of 10\arcsec\ and 
71\degr. They were chosen such that the double-cone remains small but wide enough to guarantee 
predominantly red-shifted emission in the southern intersection region with PA=35. The velocities were 
again computed under the assumption of a purely radial outflow along the cone surfaces.

The projection to the plane of the sky and the velocities along the slit positions are shown in the
middle panels of Figure \ref{fig:cone}. As before, the northern cone points away and the southern one 
towards the observer. The symmetric intersections with the slit positions and the chosen constant 
outflow velocity result in symmetric patterns of constant values in both red- and blue-shifted 
velocities. Also with this model some of the observed features might be reproduced.

Finally, we combine these two double-cones and present their projection and velocities in the
bottom panels of Figure \ref{fig:cone}. Obviously, various observed features might be better 
reproduced with these combined double-cones, while others still cannot. This example shows that the 
real physical scenario is much more complex than what can be explained by our simplistic model.
However, without further constraints from observations it will not be possible to reconstruct a 
reliable scenario.

\section{Conclusions}

We performed a multi-wavelength study of the evolved Galactic B[e] star MWC\,137 and its gaseous 
environment on both large and small scales. Measuring the kinematics of the forbidden lines in the 
optical nebula results in the general trend that the northern part is approaching us while 
the southern region is receding. Approximating the double-ring structure apparent in the optical 
nebula by a sheared double-cone it is obvious that MWC\,137 is not residing in the center of this 
geometrical pattern. Also, the axis of the double-cone is not aligned with the jet axis. We conclude 
from this that MWC\,137 alone cannot have formed and shaped the large-scale nebula. However, a second 
double-cone structure with symmetry axis aligned with the observed jet and centered on the star's 
position might be postulated. Such a combined scenario provides many consistencies with the observed 
nebula kinematics.

Knot c of the northern jet was resolved as well, which displays sub-structure and a velocity gradient 
in agreement with the general jet behavior: The radial velocity decreases with increasing distance from 
the star. 

The large-scale distribution of the cold molecular gas and dust were also investigated. 
The molecular gas was mapped in the CO(3-2) lines, and we identified primarily two major
groupings. The first one consists of four pronounced clumps north-east, east, south, and south-west
of the optical nebula forming an arc-like structure. 
The second large-scale structure is best described by a huge and more or less straight molecular belt 
spreading from the north-east to the south-west. As the velocity within this arrangement is confined to 
a very narrow range, it is most likely aligned with the plane of the sky.
As also the kinematics at the outskirts of the 
optical nebula agree with those of the molecular gas, we conclude that the cold gas is the remnant of 
the initial molecular cloud from which the cluster has formed. Support for this scenario is provided by 
the huge molecular masses that are inconsistent with material ejected during the previous evolution of 
one (or more) stars.

The dust mapped in the near- and far-infrared surrounds the optical nebula in the same way as the cold 
molecular gas. It displays an additional arc towards the south which is aligned with the southern arc 
of the cold molecular gas but which has no optical counterpart. No dust and cold gas are seen towards 
the north and the north-west of the optical nebula.

Dust clumps are also present on much smaller scales, in the close vicinity of the star where they
reflect the light coming from the innermost region, in which a circumstellar gas disk is expected
to revolve the star on Keplerian orbits. This gas disk gives rise to the rotationally broadened
CO bands that are seen in the K-band and might also host the driving mechanism for the jet.

\acknowledgments

We thank the anonymous referee for constructive comments.  
This research made use of the NASA Astrophysics Data System (ADS) and of the 
SIMBAD database, operated at CDS, Strasbourg, France. We wish to thank Eero 
Vaher for the preparation of the python script for data cube assembling. We also thank 
Moorits Mihkel Muru, Grete-Lilijane K\"{u}ppas, and Gutnar Leede for collecting the 
NOT-ALFOSC spectra during the NOT course in November 2016 at Tuorla Observatory, 
University of Turku, Finland.
Parts of the observations obtained with the MPG 2.2m telescope were supported
by the Ministry of Education, Youth and Sports project - LG14013 (Tycho
Brahe: Supporting Ground-based Astronomical Observations).
Parts of the data presented here were obtained with ALFOSC, which is provided by the Instituto de Astrofisica de Andalucia (IAA) under a joint agreement with the University of Copenhagen and NOTSA.
MK and DHN acknowledge financial support from GA\v{C}R (grant numbers 
17-02337S and 16-05011S). The Astronomical Institute Ond\v{r}ejov is supported by the project 
RVO:67985815. TL acknowledges support from the institutional research 
funding IUT40-1 of the Estonian Ministry of Education and Research. 
This research was also supported by the European Union European 
Regional Development Fund, project ``Benefits for Estonian Society from Space 
Research and Application'' (KOMEET, 2014\,-\,2020.\,4.\,01.\,16\,-\,0029).
LC, MLA, CEC, and ND acknowledge financial support from CONICET 
(PIP 0177 and 0356) and the Universidad Nacional de La Plata (Programa de Incentivos 
G11/137 and G11/139). Financial support for International Cooperation of
the Czech Republic and Argentina (AVCR-CONICET/14/003) is acknowledged.
LC and MC thank for the support from the project CONICYT + PAI/Atracci\'on
de capital humano avanzado del extranjero (folio PAI80160057).
GM acknowledges support from CONICYT, Programa de Astronom\'ia/PCI,
FONDO ALMA 2014, Proyecto No 31140024.

\bibliographystyle{aasjournal}
\bibliography{ms}

\end{document}